\documentclass[11pt,letterpaper]{article}
\usepackage{eurosym}
\usepackage{setspace}
\usepackage{color}
\usepackage{amsmath}
\usepackage{amsfonts}
\usepackage{subcaption}
\usepackage{comment}
\usepackage{booktabs}
\usepackage{verbatim}
\usepackage{amssymb}
\usepackage{graphicx}

\providecommand{\U}[1]{\protect\rule{.1in}{.1in}}
\begin{document}

\title{Crystals of superconducting Baryonic tubes in the \\
low energy limit of QCD at finite density}
\author{Fabrizio Canfora$^{1}$, Marcela Lagos$^{2}$, Aldo Vera$^{1, 2}$ \\
$^{1}$\textit{Centro de Estudios Cient\'{\i}ficos (CECS), Casilla 1469,
Valdivia, Chile,}\\
$^{2}$\textit{Instituto de Ciencias F\'isicas y Matem\'aticas, Universidad
Austral de Chile, Valdivia, Chile.}\\
{\small canfora@cecs.cl, marcelagosf@gmail.com, aldo.vera@uach.cl}}
\maketitle

\begin{abstract}
The low energy limit of QCD admits (crystals of) superconducting Baryonic
tubes at finite density. We begin with the Maxwell-gauged Skyrme model in
(3+1)-dimensions (which is the low energy limit of QCD in the leading order
of the large \textbf{N} expansion). We construct an ansatz able to reduce
the seven coupled field equations in a sector of high Baryonic charge to
just one linear Schr\"odinger-like equation with an effective potential
(which can be computed explicitly) periodic in the two spatial directions
orthogonal to the axis of the tubes. The solutions represent ordered arrays
of Baryonic superconducting tubes as (most of) the Baryonic charge and total
energy is concentrated in the tube-shaped regions. They carry a persistent
current (which vanishes outside the tubes) even in the limit of vanishing
U(1) gauge field: such a current cannot be deformed continuously to zero as
it is tied to the topological charge. Then, we discuss the subleading
corrections in the 't Hooft expansion to the Skyrme model (called usually $%
\mathcal{L}_{6}$, $\mathcal{L}_{8}$ and so on). Remarkably, the very same
ansatz allows to construct analytically these crystals of superconducting
Baryonic tubes at any order in the 't Hooft expansion. Thus, no matter how
many subleading terms are included, these ordered arrays of gauged solitons
are described by the same ansatz and keep their main properties manifesting
a universal character. On the other hand, the subleading terms can affect
the stability properties of the configurations setting lower bounds on the
allowed Baryon density.
\end{abstract}

\newpage

\tableofcontents

\section{Introduction}

Exact analytic results on the phase diagram of the low energy limit of QCD
at finite density and low temperatures are extremely rare (it is often
implicitly assumed that they are out of reach of the available techniques).
This fact, together with the non-perturbative nature of low energy QCD, is
one of the main reasons why it is far from easy to have access to the very
complex and interesting structure of the phase diagram (see \cite{newd3}-%
\cite{newd6}, and references therein) with analytic techniques.

One of the most intriguing phenomena that arises in the QCD phase diagram at
very low temperatures and finite Baryon density, is the appearance of
ordered structures like crystals of solitons (as it happens, for instance,
in condensed matter theory with the Larkin--Ovchinnikov--Fulde--Ferrell
phase \cite{LOFF1yLOFF2}). From the numerical and phenomenological point of
view, ordered structures are expected to appear in the low energy limit of
QCD (see, for instance, \cite{newd7}-\cite{Zahed}, and references therein).
The available analytic results have been found in $(1+1)$-dimensional toy
models and all of them suggest the appearance of ordered structures\footnote{%
The results in \cite{newd13y16} clearly show that, quite generically in $%
(1+1)$-dimensions, there is a phase transition at a critical temperature
from a massless phase to a broken phase with a non-homogeneous condensate as
it also happens in superconductors \cite{LOFF1yLOFF2}.} of solitons.

Even less is known when the electromagnetic interactions arising within
these ordered structures are turned on. Analytic examples of crystals of
gauged solitons with high topological charge in $(3+1)$-dimensions in the
low energy limit of QCD would reveal important physical aspects of these
ordered phases. The only available analytical examples\footnote{%
See \cite{Ward} for the construction of (quasi-)periodic non-trivial
solutions in one spatial direction (Skyrme chains) using approximate
analytical methods.} are derived either in lower dimensions and/or when some
extra symmetries (such as SUSY) are included (see \cite{31a40}, and
references therein).

We search for analytic solutions (despite the fact that these questions can
be addressed numerically, as the previous references show) because a
systematic tool to construct analytic crystals of gauged solitons can
greatly enlarge our understanding of the low energy limit of QCD: the
analytic tools developed here below disclose novel and unexpected phenomena.

The gauged Skyrme model \cite{skyrme}, \cite{Witten}, which describes the
low energy limit of QCD minimally coupled with Maxwell theory at the leading
order in the 't Hooft expansion \cite{witten0}-\cite{ANW} (for two detailed
reviews see \cite{manton} and \cite{BaMa}), will be our starting point.
Using the methods introduced in \cite{56b}, \cite{56}, \cite{gaugsk} and 
\cite{gaugsksu(n)} we will construct analytic gauged multi-soliton solutions
at finite Baryon density with crystalline structure and high topological
charge. These crystals describe ordered arrays of superconducting tubes in
which (most of) the topological charge and total energy are concentrated
within tube-shaped regions\footnote{%
In \cite{Jackson} and \cite{Nitta:2007zv}, numerical string solutions in the
Skyrme model with mass term have been constructed. However, those
configurations are classically unstable because they have a zero topological
density (then are expected to decay into Pions). The new solutions
constructed in the present paper are topologically protected and therefore
can not decay in those of \cite{Jackson}, \cite{Nitta:2007zv}.}. They carry
a persistent current (which vanishes outside the tubes) which cannot be
deformed continuously to zero as it is tied to the topological charge.

These regular superconducting tubes can be considered as explicit analytic
examples of the superconducting strings introduced in \cite{wittenstrings}.
The spectacular observable effects that such objects could have (see \cite%
{shifman1aimportanceshifman}, and references therein) together with the fact
that these objects can be constructed using natural ingredients generate a
huge interest both theoretically and phenomenologically. However, until now,
there are very few explicit analytic $(3+1)$-dimensional examples built
using only ingredients arising from the standard model. In fact, the present
superconducting tubes appear in the low energy limit of QCD minimally
coupled with Maxwell theory.

Then we move to the subleading correction to the Skyrme model in the 't
Hooft expansion (see \cite{marleau1}-\cite{Gudnason:2017opo} and references
therein). Although one could believe that such complicated corrections could
destroy the nice analytic properties of the crystals of superconductive
Baryonic tubes, we will show that \textit{no matter how many subleading
terms are included, these ordered arrays of gauged solitons are described by
the very same ansatz and keep unchanged their main properties manifesting a
clear universal character}. On the other hand, the stability properties of
these crystals of superconducting Baryonic tubes can be affected in a
non-trivial way by the subleading terms in the 't Hooft expansion.

The paper is organized as follows. In the second section the general field
equations will be derived and the definition of topological charge will be
introduced. In the third section, the ansatz which allows to solve
analytically the field equations (no matter how many subleading terms are
included) in the ungauged case in a sector with high topological charge will
be discussed. Then, it will be explained how this ansatz can be generalized
to the gauged case with the inclusion of the minimal coupling with the
Maxwell field. The physical properties of these gauged crystals and their
universal character will be analyzed. We will also study how the subleading
terms can affect the stability properties of the configurations setting
lower bounds on the allowed Baryon density. In the last sections, some
conclusions and perspectives will be presented.

\section{The gauged generalized Skyrme model}

The starting point is the action of the U(1) gauged Skyrme model in four
dimensions, which corresponds to the low energy limit of QCD at leading
order in the 't Hooft expansion: 
\begin{gather}
I=\int \frac{d^{4}v}{4}\left[ K\mathrm{Tr}\left( L^{\mu }L_{\mu }+\frac{%
\lambda }{8}G_{\mu \nu }G^{\mu \nu }\right) -\left( 2m\right) ^{2}\mathrm{Tr}%
\left( U+U^{-1}\right) -F_{\mu \nu }F^{\mu \nu }+\mathcal{L}_{\text{corr}}%
\right] \ ,  \label{sky1} \\
L_{\mu }=U^{-1}D_{\mu }U\ ,\ \ G_{\mu \nu }=[L_{\mu },L_{\nu }]\ ,\ \ D_{\mu
}=\nabla _{\mu }+A_{\mu }\left[ t_{3},\ .\ \right] \ ,\ d^{4}v=d^{4}x\sqrt{-g%
}\ ,  \label{sky2} \\
U\in \text{SU(2)}\ ,\ \ L_{\mu }=L_{\mu }^{j}t_{j}\ ,\ \ t_{j}=i\sigma _{j}\
,\ F_{\mu \nu }=\partial _{\mu }A_{\nu }-\partial _{\nu }A_{\mu }\ ,
\label{sky2.5}
\end{gather}%
where $K$ and $\lambda \ $are the Skyrme couplings, $d^{4}v$ is the
four-dimensional volume element, $g$ is the metric determinant, $m$ is the
Pions mass, $A_{\mu }$ is the gauge potential, $\nabla _{\mu }$ is the
partial derivative and $\sigma _{i}$ are the Pauli matrices. In Eq. (\ref%
{sky1}), $\mathcal{L}_{\text{corr}}$ represents the possible subleading
corrections to the Skyrme model which can be computed, in principle, using
either Chiral Perturbation Theory (see \cite{Scherer:2002tk} and references
therein) or the large \textbf{N} expansion \cite{tHooft:1973alw}, \cite%
{Witten:1979kh}. The expected corrections have the following generic form 
\begin{align}
\mathcal{L}_{6}=& \frac{c_{6}}{96}\text{Tr}\left[ G_{\mu }{}^{\nu }G_{\nu
}{}^{\rho }G_{\rho }{}^{\mu }\right] \ ,  \notag \\
\mathcal{L}_{8}=& -\frac{c_{8}}{256}\biggl(\text{Tr}\left[ G_{\mu }{}^{\nu
}G_{\nu }{}^{\rho }G_{\rho }{}^{\sigma }G_{\sigma }{}^{\mu }\right] -\text{Tr%
}\left[ \{G_{\mu }{}^{\nu },G_{\rho }{}^{\sigma }\}G_{\nu }{}^{\rho
}G_{\sigma }{}^{\mu }\right] \biggl)\ ,  \label{Lcorr}
\end{align}%
and so on \cite{marleau1}, where the $c_{p}$ ($p\geq 6$) are subleading with
respect to $K$ and $\lambda $.

A natural question arises here: \textit{in which sense these correction
terms are subleading with respect to the original Skyrme model?} First of
all, we would like to remark that this question does not affect directly the
present construction since the analytic method presented here allows to
construct exact solutions no matter how many \textquotedblleft subleading
terms\textquotedblright\ are included (as it will be shown in the following
sections). However, from the physical point of view, the above question is
very interesting. In principle, as remarked in \cite{AdkinsNappi}, \cite%
{JacksonJackson} and \cite{Gudnason:2017opo}, one should expect generically
higher-derivative terms of the chiral field $U$ in the low-energy limit of
QCD. Due to the fact that each term is larger in canonical dimension,
dimensional constants to the same power minus four must go with each of
them. These constants are expected to be proportional to the mass scale of
the degrees of freedom integrated out of the underlying theory. Therefore,
as long as the energy scales being probed are much smaller than the lowest
mass scale of a state that was integrated out, the higher-derivative
expansion may make sense and thus converge. Another intuitive argument is
due to `t Hooft and Witten (in the classic references \cite{witten0}, \cite%
{tHooft:1973alw} and \cite{Witten:1979kh}) which argued that in the large-%
\textbf{N} limit, QCD becomes equivalent to an effective field theory of
mesons and which the higher order terms with respect to the non-linear sigma
model (NLSM henceforth) action are accompanied by inverse power of \textbf{N}
(where here \textbf{N} is the number of colors).

The analysis here below will clarify that no matter how many further
subleading terms are included, the pattern will never change. In particular,
using precisely the same ansatz discussed in the next section for the SU(2)
-valued Skyrmionic field, the field equations for the generalized Skyrme
model with all the corrections included always reduce to a first order
integrable\footnote{%
Here \textquotedblleft integrable" has the usual meaning: an integrable
differential equation is an equation which can be reduced to a quadrature.
This important property allows to reduce the computation of the total energy
(as well as of other relevant physical properties) to definite integrals of
elementary functions.} ordinary differential equation.

Moreover, if the minimal coupling with Maxwell theory is included, the
gauged version of the field equations for the SU(2)-valued Skyrmionic field
remains explicitly integrable (namely, despite the minimal coupling with the
Maxwell field, the soliton profile can still be determined analytically)
while the four Maxwell equations with the U(1) current arising from the
minimal coupling with the generalized Skyrme model (with all the subleading
terms included) reduce to a single linear Schr\"odinger-like equation for
the relevant component of the Maxwell potential in which the effective
potential can be computed explicitly in terms of the solitons profile.

It is worth to emphasize that this results is quite remarkable: not only the
three non-linear SU(2) coupled field equations in the ungauged case
(including all the subleading corrections to the Skyrme model) can always be
reduced to a single integrable first order ODE in a sector with arbitrary
Baryonic charge in (3+1)-dimensions. Moreover, in the gauged case minimally
coupled with the Maxwell theory, the fully coupled seven non-linear field
equations (three from the generalized Skyrme model plus the four Maxwell
equations with the corresponding current) are reduced to the very same
integrable ODE for the profile plus a linear Schr\"odinger-like equation for
the relevant component of the Maxwell field in which the effective potential
can be computed explicitly in terms of the solitons profile itself. Without
such a reduction, even the numerical analysis of the electromagnetic
properties of these (3+1)-dimensional crystals of superconducting tubes
would be a really hard task (which up to now, has not been completed to the
best of our knowledge). While, with the present approach, the numerical task
to analyze the electromagnetic properties of these crystal of
superconducting tubes is reduced to a linear Schr\"odinger equation with an
explicitly known potential.

\subsection{Field equations}

The field equations of the model are obtained varying the action in Eq. %
\eqref{sky1} w.r.t. the $U$ field and the Maxwell potential $A_\mu$. To
perform this derivation it is useful to keep in mind the following relations 
\begin{gather*}
\delta_U L_{\mu }=[L_{\mu },U^{-1}\delta U]+D _{\mu }(U^{-1}\delta U) \ , \\
\delta_U G_{\mu \nu }=D_{\nu }[L_{\mu },U^{-1}\delta U]-D_{\mu}[L_{\nu
},U^{-1}\delta U]\ ,
\end{gather*}%
where $\delta_U$ denotes variation w.r.t the $U$ field, and 
\begin{align*}
& \frac{\delta }{\delta A^{\mu }}\biggl(\text{Tr}(L_{\nu }L^{\nu })\biggl)\
=\ 2\text{Tr}(\hat{O}L_{\mu })\ , \\
& \frac{\delta }{\delta A^{\mu }}\biggl(\text{Tr}(G_{\alpha \beta }G^{\alpha
\beta })\biggl)\ =\ 4\text{Tr}\biggl(\hat{O}[L^{\nu },G_{\mu \nu }]\biggl)\ ,
\\
& \frac{\delta }{\delta A^{\mu }}\biggl(\text{Tr}(G_{\alpha }{}^{\nu }G_{\nu
}{}^{\rho }G_{\rho }{}^{\alpha })\biggl)\ =\ 3\text{Tr}\biggl(\hat{O}%
[L^{\alpha },[G_{\mu \nu },G_{\alpha }{}^{\nu }]]\biggl)\ , \\
& \frac{\delta }{\delta A^{\mu }}\biggl(\text{Tr}(G_{\alpha }{}^{\nu }G_{\nu
}{}^{\rho }G_{\rho }{}^{\sigma }G_{\sigma }{}^{\alpha })\biggl)\ =\ 4\text{Tr%
}\biggl(\hat{O}[L^{\alpha },G_{\alpha }{}^{\nu }G_{\nu }{}^{\rho }G_{\rho
\mu }-G_{\mu}{}^\nu G_{\nu}{}^\rho G_{\rho\alpha}]\biggl)\ , \\
& \frac{\delta }{\delta A^{\mu }}\biggl(\text{Tr}(\{G_{\alpha }{}^{\nu
},G_{\rho }{}^{\sigma }\}G_{\nu }{}^{\rho }G_{\sigma }{}^{\alpha })\biggl)\
=\ 4\text{Tr}\biggl(\hat{O}[\{G_{\nu }{}^{\rho },\{G_{\mu }{}^{\nu },G_{\rho
}{}^{\sigma }\}\},L_{\sigma }]\biggl)\ .
\end{align*}%
Here we have used 
\begin{equation*}
\frac{\delta G_{\beta }{}^{\alpha }}{\delta A^{\mu }}=\delta _{\mu \beta }[%
\hat{O},L^{\alpha }]+\delta _{\mu }^{\alpha }[L_{\beta },\hat{O}] \ ,
\end{equation*}
and we have defined 
\begin{equation*}
\frac{\delta L_{\nu }}{\delta A^{\mu }}=\delta _{\mu \nu }\hat{O}\ ,\qquad 
\hat{O}\ =\ U^{-1}[t_{3},U]\ .
\end{equation*}
From the above, the field equations of the gauged generalized Skyrme model
turns out to be 
\begin{align}
\frac{K}{2}\biggl(D^{\mu }L_{\mu }+\frac{\lambda }{4}D^{\mu }[L^{\nu},G_{\mu
\nu }]\biggl)+2m^{2}\left( U-U^{-1}\right) +3c_{6}[L_{\mu },D_{\nu }[G^{\rho
\nu },G_{\rho }{}^{\mu }]] &  \notag \\
+4c_{8}\biggl[L_{\mu },D_{\nu }\biggl(G^{\nu \rho }G_{\rho \sigma }G^{\sigma
\mu}+G^{\mu \rho }G_{\rho \sigma }G^{\nu \sigma }+\{G_{\rho \sigma
},\{G^{\mu \rho },G^{\nu \sigma }\}\}\biggl)\biggl] & \ = \ 0\ ,
\label{sessea0}
\end{align}%
together with 
\begin{equation}
\nabla _{\mu }F^{\mu \nu }=J^{\nu }\ ,
\end{equation}%
where the current $J_\mu$ is given by 
\begin{align}
J_{\mu } \ = \ & \frac{K}{2}\text{Tr}\biggl[\hat{O}\biggl(L_{\mu }+\frac{%
\lambda }{4}[L^{\nu },G_{\mu \nu }]\biggl)\biggl]+\frac{c_{6}}{32}\text{Tr}%
\biggl[\hat{O}\biggl(\lbrack L^{\alpha },[G_{\mu \nu },G_{\alpha }{}^{\nu }]]%
\biggl)\biggl]  \notag \\
& -\frac{c_{8}}{64}\text{Tr}\biggl[\hat{O}\biggl(\lbrack L^{\alpha
},G_{\alpha }{}^{\nu }G_{\nu }{}^{\rho }G_{\rho \mu }+G_{\rho \mu }G_{\nu
}{}^{\rho }G_{\alpha}{}^{\nu }+\{G^{\nu\rho },\{G_{\mu \nu },G_{\rho\alpha
}\}\}]\biggl)\biggl]\ .  \label{Jmu}
\end{align}

\subsection{Energy-momentum tensor and topological charge}

Using the standard definition 
\begin{equation}  \label{HilbertTmunu}
T_{\mu \nu }\ =\ -2\frac{\partial \mathcal{L}}{\partial g^{\mu \nu }}+g_{\mu
\nu }\mathcal{L}\ ,
\end{equation}%
we can compute the energy-momentum tensor of the theory under consideration 
\begin{equation}  \label{Tmunu}
T_{\mu \nu }\ =\ T_{\mu \nu }^{\text{Sk}}+T_{\mu \nu }^{\text{mass}}+T_{\mu
\nu }^{(6)}+T_{\mu \nu }^{(8)}+T_{\mu \nu }^{\text{U(1)}}\ ,
\end{equation}%
with $T_{\mu \nu }^{\text{U(1)}}$ the energy-momentum tensor of the Maxwell
field 
\begin{align*}
T_{\mu\nu}^{\text{\text{U(1)}}} = F_{\mu\alpha}F_{\nu}^{\;\alpha}-\frac{1}{4}%
g_{\mu\nu} F_{\alpha\beta}F^{\alpha\beta} \ .
\end{align*}
According to Eq. \eqref{HilbertTmunu}, a direct computation reveals that 
\begin{align*}
T_{\mu \nu }^{\text{mass}}=& -m^{2}g_{\mu \nu }\text{Tr}(U+U^{-1})\ , \\
T_{\mu \nu }^{\text{Sk}}=& -\frac{K}{2}\text{Tr}\biggl(L_{\mu }L_{\nu }-%
\frac{1}{2}g_{\mu \nu }L_{\alpha }L^{\alpha }+\frac{\lambda }{4}(g^{\alpha
\beta }G_{\mu \alpha }G_{\nu \beta }-\frac{1}{4}g_{\mu \nu }G_{\alpha \beta
}G^{\alpha \beta })\biggl)\ , \\
T_{\mu \nu }^{(6)}=& -\frac{c_{6}}{16}\text{Tr}\biggl(g^{\alpha \gamma
}g^{\beta \rho }G_{\mu \alpha }G_{\nu \beta }G_{\gamma \rho }-\frac{1}{6}%
g_{\mu \nu }G_{\alpha }{}^{\beta }G_{\beta }{}^{\rho }G_{\rho }{}^{\alpha }%
\biggl)\ , \\
T_{\mu \nu }^{(8)}=& \ \frac{c_{8}}{32}\text{Tr}\biggl(g^{\alpha \rho
}g^{\beta \gamma }g^{\delta \lambda }G_{\alpha \mu }G_{\nu \beta }G_{\gamma
\delta }G_{\lambda \rho }+\frac{1}{2}\{G_{\mu \alpha },G_{\lambda \rho
}\}\{G_{\beta \nu },G_{\gamma \delta }\}g^{\alpha \gamma }g^{\beta \rho
}g^{\delta \lambda }  \notag \\
& -\frac{1}{8}g_{\mu \nu }(G_{\alpha }{}^{\beta }G_{\beta }{}^{\rho }G_{\rho
}{}^{\sigma }G_{\sigma }{}^{\alpha }-\{G_{\alpha }{}^{\beta },G_{\rho
}{}^{\sigma }\}G_{\beta }{}^{\rho }G_{\sigma }{}^{\alpha })\biggl)\ .
\end{align*}

The topological charge of the gauged Skyrme model is given by \cite{Witten}, 
\cite{gaugesky1}: 
\begin{gather}  \label{new4.1.1}
B=\frac{1}{24\pi^{2}}\int_{\Sigma}\rho_{\text{B}}\ , \\
\rho_{\text{B}}=\epsilon^{ijk}\text{Tr}\biggl[\left(
U^{-1}\partial_{i}U\right) \left( U^{-1}\partial_{j}U\right) \left(
U^{-1}\partial_{k}U\right) -\partial_{i}\left[ 3A_{j}t_{3}\left(
U^{-1}\partial_{k}U+\left( \partial_{k}U\right) U^{-1}\right) \right] \biggl]%
\ .
\end{gather}
Note that the second term in Eq. \eqref{new4.1.1}, the Callan-Witten term,
guarantees both the conservation and the gauge invariance of the topological
charge. When $\Sigma$ is space-like, $B$ is the Baryon charge of the
configuration.

\section{Crystals of superconducting Baryonic tubes}

In this section we will show that the gauged generalized Skyrme model admits
analytical solutions describing crystals of superconducting Baryonic tubes
at finite density.

\subsection{The ansatz}

Finite density effects can be accounted for using the flat metric defined
below: 
\begin{equation}
ds^{2}=-dt^{2}+L^{2}\left( dr^{2}+d\theta^{2}+d\phi^{2}\right) \ ,
\label{Minkowski}
\end{equation}
where $4\pi^{3}L^{3}$ is the volume of the box in which the gauged solitons
are living. The adimensional coordinates have the ranges $0\leq r\leq2\pi\ $%
, $0\leq\theta\leq\pi\ $, $0\leq\phi\leq2\pi$.

For the Skyrme field we use the standard SU(2) parameterization 
\begin{gather}
U^{\pm 1}(x^{\mu })=\cos \left( \alpha \right) \mathbf{1}_{2}\pm \sin \left(
\alpha \right) n^{i}t_{i}\ ,\ \ n^{i}n_{i}=1\ ,  \label{sessea1} \\
n^{1}=\sin \Theta \cos \Phi \ ,\ \ n^{2}=\sin \Theta \sin \Phi \ ,\ \
n^{3}=\cos \Theta \ .  \label{sessea2}
\end{gather}%
From Eqs. \eqref{sessea1} and \eqref{sessea2} the topological charge density
reads 
\begin{equation*}
\rho _{\text{B}}=-12(\sin ^{2}{\alpha }\sin{\Theta })d\alpha \wedge d\Theta
\wedge d\Phi \ ,
\end{equation*}%
and therefore, as we want to consider only topologically non-trivial
configurations, we must demand that 
\begin{equation}
d\alpha \wedge d\Theta \wedge d\Phi \neq 0\ .  \label{sessea3}
\end{equation}%
Now, the problem is to find a good ansatz which respect the above condition
and simplify as much as possible the field equations. A close look at Eq. (%
\ref{sessea0}) (see Appendix II for its explicit form in terms of $\alpha$, $%
\Theta$ and $\Phi$) reveals that a good set of conditions is 
\begin{equation}
\nabla _{\mu }\Phi \nabla ^{\mu }\alpha =\nabla _{\mu }\alpha \nabla ^{\mu
}\Theta =\nabla _{\mu }\Phi \nabla ^{\mu }\Phi =\nabla _{\mu }\Theta \nabla
^{\mu }\Phi =0\ .  \label{Appconds}
\end{equation}%
A suitable choice that satisfies Eqs. (\ref{sessea3}) and (\ref{Appconds})
is the following \cite{gaugsk}: 
\begin{equation}
\alpha =\alpha \left( r\right) \ , \quad \Theta =q\theta \ , \quad \Phi
=p\left( \frac{t}{L}-\phi \right) \ , \quad q=2v+1\ , \quad p,v\in \mathbb{N}%
\ ,\ p\neq 0\ .  \label{good1}
\end{equation}%
Additionally some other useful relations are satisfied by the above ansatz,
namely 
\begin{equation}
\Box \Theta =\Box \Phi =0\ .  \label{good2}
\end{equation}

\subsection{Solving the system analytically}

The identities in Eqs. (\ref{Appconds}) and (\ref{good2}) satisfied by the
ansatz in Eq. (\ref{good1}) greatly simplify the field equations keeping
alive the topological charge. This can be seen as follows.

\textit{Firstly}, a direct inspection of the field equations reveals that
all the terms which involve $\sin ^{2}\Theta $ are always multiplied by $%
\nabla _{\mu }\Phi \nabla ^{\mu }\Phi $ so that all such terms disappear.

\textit{Secondly}, since $\Theta $ is a linear function, in all the terms in
the field equations $\nabla _{\mu }\Theta \nabla ^{\mu }\Theta $ becomes
just a constant.

\textit{Thirdly}, since the gradients of $\alpha $, $\Theta $\ and $\Phi $\
are mutually orthogonal (and, moreover, $\nabla ^{\mu }\Phi $\ is a
light-like vector), all the terms in the field equations which involve $%
\nabla _{\mu }\Phi \nabla ^{\mu }\alpha $, $\nabla _{\mu }\alpha \nabla
^{\mu }\Theta $, $\nabla _{\mu }\Phi \nabla ^{\mu }\Phi $ and $\nabla _{\mu
}\Theta \nabla ^{\mu }\Phi $ vanish.

\textit{Fourthly}, the above three properties together with Eq. (\ref{good2}%
) ensures that two of the three field equations of the generalized Skyrme
model are identically satisfied (see Appendix I and Appendix II).

It is also worth to emphasize that the four properties listed here above are
true no matter how many subleading terms ($\mathcal{L}_{10}$, $\mathcal{L}%
_{12}$ and so on) are included in the generalized Skyrme action. For the
above reasons, the three non-linear coupled field equations of the
generalized Skyrme model in Eq. (\ref{sessea0}) with the ansatz in Eq. (\ref%
{good1}) are reduced to the following single ODE for the profile\footnote{%
It is interesting to note that the terms in the field equations arising from 
$\mathcal{L}_{6}$ in the generalized Skyrme model vanish identically due to
the properties of the ansatz in Eqs. (\ref{Appconds}), (\ref{good1}) and (%
\ref{good2}). On the other hand, such a term can affect the stability
properties of the solutions, as we will see below.} $\alpha $: 
\begin{align}
\alpha ^{\prime \prime }-\frac{q^{2}}{2}\sin (2\alpha )+\frac{4m^{2}L^{2}}{K}%
\sin (\alpha )+\frac{\lambda q^{2}}{L^{2}}\sin (\alpha )[\cos (\alpha
)\alpha ^{\prime 2}+\sin (\alpha )\alpha ^{\prime \prime }]&   \notag \\
-\frac{3c_{8}q^{4}}{KL^{6}}\sin ^{3}(\alpha )[\cos (\alpha )\alpha ^{\prime
2}+\sin (\alpha )\alpha ^{\prime \prime }]\alpha ^{\prime 2}& \ =\ 0\ .
\label{sessea4}
\end{align}%
This is already a quite interesting fact in itself. Moreover, the above
analysis clearly shows that it will remain true even including further
subleading term. What is really remarkable is that Eq. (\ref{sessea4}) can
always be reduced to a first order ODE: 
\begin{equation}
\left\{ Y\left( \alpha \right) \left( \alpha ^{\prime }\right) ^{2}+W\left(
\alpha \right) +E_{0}\right\} ^{\prime }=0\ ,  \label{sessea5}
\end{equation}%
\begin{equation*}
Y\left( \alpha \right) =1+\frac{\lambda q^{2}}{L^{2}}\sin ^{2}(\alpha )-%
\frac{3c_{8}q^{4}}{2KL^{6}}\sin ^{4}(\alpha )\alpha ^{\prime 2}\ ,\qquad
W(\alpha )=\frac{q^{2}}{2}\cos (2\alpha )-\frac{8m^{2}L^{2}}{K}\cos (\alpha
)\ ,
\end{equation*}%
which is explicitly solvable in terms of generalized Elliptic Integrals \cite%
{Elliptics}. Here $E_{0}$ is a positive integration constant and $X^{\prime
}=\frac{dX}{dr}$. Therefore Eq. (\ref{sessea5}) implies that, with the
ansatz defined in Eq. (\ref{good1}), the field equations are integrable and
reducible to the following quadrature\footnote{%
The identities in Eqs. (\ref{Appconds}) and (\ref{good2}) satisfied by the
ansatz in Eq. (\ref{good1}) ensures that (even if the subleading corrections 
$\mathcal{L}_{10}$, $\mathcal{L}_{12}$ and so on are included) the ansatz in
Eq. (\ref{good1}) will always reduce the three coupled non-linear field
equations to a single first order ODE for the profile $\alpha $.}:%
\begin{gather}
\frac{d\alpha }{\chi (\alpha ,E_{0})}=\pm dr\ ,  \label{sessea5.2} \\
\chi (\alpha ,E_{0})=\pm \sqrt{\frac{KL^{6}\csc ^{4}{(\alpha )}}{3c_{8}q^{4}}%
\biggl(1+\frac{q^{2}\lambda }{L^{2}}\sin ^{2}{(\alpha )}+\sqrt{\frac{%
6c_{8}q^{4}}{KL^{6}}(W+E_{0})\sin ^{4}{(\alpha )}+\biggl(1+\frac{%
q^{2}\lambda }{L^{2}}\sin ^{2}{(\alpha )}\biggl)^{2}}\biggl)}\ ,
\label{sessea5.2.1}
\end{gather}%
with $E_{0}\geq -W-\frac{KL^{6}}{3c_{8}q^{4}\sin ^{4}(\alpha )}(1+\frac{%
q^{2}\lambda }{L^{2}}\sin ^{2}(\alpha ))^{2}$, for $c_{8}>0$. It is also
worth to emphasize that the integration constant $E_{0}$ can be chosen in
such a way that, first of all, $\alpha ^{\prime }$\ never changes sign
(which is a necessary condition for stability) and, secondly, the
topological charge is $B=np$ (as we will show in the following subsection).

Quite surprisingly, these very intriguing properties of the ansatz are not
destroyed by the inclusion of the minimal coupling with Maxwell field. The
coupling of the generalized Skyrme model with the Maxwell theory is
introduced replacing the partial derivatives acting on the SU(2)-valued
scalar field $U$ with the following covariant derivative 
\begin{equation}
\nabla_{\mu }U\rightarrow D_{\mu }U=\nabla_{\mu }U+A_{\mu }[t_{3},U]\ .
\label{covdev1}
\end{equation}%
A straightforward computation shows that the above replacement in Eq. (\ref%
{covdev1}) is completely equivalent to the replacement here below (in terms
of $\alpha $, $\Theta $\ and $\Phi $)%
\begin{gather}
\nabla_{\mu }\alpha \rightarrow \nabla_{\mu }\alpha \ ,\quad \nabla _{\mu
}\Theta \rightarrow \nabla_{\mu }\Theta \ ,\quad \nabla_{\mu }\Phi
\rightarrow D_{\mu }\Phi =\nabla_{\mu }\Phi -2A_{\mu }\Phi \ .
\label{covdev2}
\end{gather}%
It is worth to emphasize that $D_{\mu }\Phi $ determines the
\textquotedblleft direction'' of the electromagnetic current (as it will be
discussed below).

Obviously, when the derivative is replaced with the Maxwell covariant
derivative (as defined in Eq. (\ref{covdev1}) or, equivalently, in Eq. (\ref%
{covdev2})), in the field equations of the gauged generalized Skyrme model
many new terms will appear which couple the SU(2) degrees of freedom with
the U(1) gauge potential $A_{\mu }$. Thus, one may ask:

\textit{Which is the best choice of the ansatz for the gauge potential} $%
A_{\mu }$ \textit{which keeps as much as possible the very nice properties
of the ansatz of the} SU(2)-\textit{valued scalar field in Eqs. (\ref%
{sessea3}), (\ref{Appconds}) and (\ref{good2}) which allowed the complete
analytic solutions in the previous case}?

In order to achieve this goal, it is enough to demand 
\begin{gather}
\nabla ^{\mu }A_{\mu }=0\ ,\ A_{\mu }A^{\mu }=0\ ,\ A_{\mu }\nabla ^{\mu
}\Phi =0\ ,  \label{good4} \\
A_{\mu }\nabla ^{\mu }\alpha =0\ ,\ A_{\mu }\nabla ^{\mu }\Theta =0\ .
\label{good5}
\end{gather}%
The above conditions determine that the Maxwell potential $A_{\mu }$ must be
of the form \cite{gaugsk}: 
\begin{equation}
A_{\mu }=(u(r,\theta ),0,0,-Lu(r,\theta ))\ .  \label{good6}
\end{equation}%
From the expressions of $L_{\mu }$ (see Appendix I) one can see that,
despite the explicit presence of $A_{\mu }$ in the U(1)-covariant
derivative, the three field equations of the gauged generalized Skyrme model
still reduce to the Eq. (\ref{sessea4}). The reason is that all the
potential terms which, in principle, could couple the SU(2)-valued scalar
field $U$ with $A_{\mu }$ in the field equations actually vanish due to the
identities in Eqs. (\ref{Appconds}), (\ref{good2}), (\ref{good4}) and (\ref%
{good5}) satisfied by the choice of our ansatz (that is why we have chosen
the ansatz in that way). One can verify easily that the four Maxwell
equations are reduced to the following single PDE: 
\begin{gather}
\partial _{r}^{2}u+\partial _{\theta }^{2}u+\frac{2}{L^{2}}\sin ^{2}(\alpha
)\sin ^{2}(q\theta )\Omega (\alpha )\left( 2u-\frac{p}{L}\right) \ = \ 0 \ ,
\label{preMax}
\end{gather}%
where $\Omega (\alpha )$ is given by 
\begin{gather}
\Omega (\alpha )=KL^{2}(L^{2}+q^{2}\lambda \sin ^{2}(\alpha )+\lambda \alpha
^{\prime 2})+q^{2}\sin ^{2}(\alpha )\alpha ^{\prime 2}\biggl[2c_{6}+\frac{%
c_{8}}{L^{2}}(q^{2}\sin ^{2}(\alpha )+\alpha ^{\prime 2})\biggl]\ .
\label{Omegadef}
\end{gather}%
Note also that Eq. \eqref{preMax} can be written as a periodic
two-dimensional Schr\"odinger equation 
\begin{gather}
\Delta \Psi +V\Psi \ =\ 0\ ,\qquad \Delta =\partial _{r}^{2}+\partial
_{\theta }^{2}\ ,  \label{sesseanewmax} \\
\Psi \ =\ \frac{2L}{p}u-1\ ,\qquad V=\frac{4}{L^{2}}\sin ^{2}(\alpha )\sin
^{2}(q\theta )\Omega (\alpha )\ .  \label{Vdef}
\end{gather}%
Therefore, with the ansatz defined in Eqs. \eqref{Minkowski}, \eqref{good1}
and \eqref{good6} the seven coupled field equations of the gauged
generalized  Skyrme model minimally coupled with Maxwell theory reduce
consistently to just one linear Schr\"odinger-like equation in which the
effective two-dimensional periodic potential can be computed explicitly.
Also, the integrability of the field equations is not spoiled by any of the
subleading corrections parameterized by the $c_{p}$. Moreover, due to the
presence of quadratic and higher order terms in $A_{\mu }$ in the gauged
generalized Skyrme model which couple $A_{\mu }$ with the SU(2)-valued
scalar field $U$ (as it happens in the Ginzburg-Landau description of
superconductors), even when $A_{\mu }=0$, the current does not vanish. Such
a residual current cannot be deformed continuously to zero, and the reason
is that the only way to \textquotedblleft kill'' it would also kill the
topological charge but, as it is well known, there is no continuous
transformation which can change the topological charge. We will return to
this very important issue when we address the superconducting nature of the
Baryonic tubes.

\subsection{Boundary conditions and Baryonic charge}

From Eq. \eqref{new4.1.1} we can compute the energy density of the
configurations presented above, which turns out to be 
\begin{gather*}
\rho _{\text{B}}^{\text{tot}}=\rho _{\text{B}}+\rho _{\text{B}}^{\text{%
Maxwell}}\ , \\
\rho _{\text{B}}=-12pq\sin (q\theta )\sin ^{2}(\alpha )\partial _{r}\alpha \
, \\
\rho _{\text{B}}^{\text{Maxwell}}=12L\biggl[\biggl(2q\sin (q\theta )\sin
^{2}(\alpha )u-\cos (q\theta )\partial _{\theta }u\biggl)\partial _{r}\alpha
-q\sin (\alpha )\cos (\alpha )\sin (q\theta )\partial _{r}u\biggl]\ .
\end{gather*}%
The above expression can be written conveniently as 
\begin{equation*}
\rho _{\text{B}}^{\text{tot}}=3q\ \frac{\partial }{\partial r}\Big(p\sin
(q\theta )\ \big(\sin (2\alpha )-2\alpha \big)-2L\sin (q\theta )u\sin
(2\alpha )\Big)-\frac{\partial }{\partial \theta }\big(12L\alpha ^{\prime
}u\cos (q\theta )\big)\ ,
\end{equation*}%
and one can check that the topological charge becomes 
\begin{equation*}
B=-np-\frac{L}{\pi }\int_{0}^{2\pi }dr\ \alpha ^{\prime }\Big((-1)^{q}\
u(r,\pi )-u(r,0)\Big)\ .
\end{equation*}%
Assuming the following boundary condition for $u$ and $\alpha $ 
\begin{equation}
u(r,\pi )=(-1)^{q}\ u(r,0)\ ,\qquad \alpha \left( 2\pi \right) -\alpha
\left( 0\right) =n\pi \ ,  \label{bc1}
\end{equation}%
and taking into account that $q$ is an odd integer, the topological charge
becomes 
\begin{equation*}
B\ =\ -np\ .
\end{equation*}
It is worth to stress here that (unlike what happens in the case of a single
Skyrmion in a flat space without boundaries, when the boundary conditions
are just dictated by the condition to have finite energy) when finite
density/volume effects are taken into account the choice of the boundary
conditions is not unique anymore. A very important requirement that any
reasonable choice of boundary conditions must satisfy is that the integral
of the topological density (which, of course, by definition is the
topological charge itself) over the volume occupied by the solutions must be
an integer. If this condition is not satisfied, the configurations would not
be well defined. Hence, the boundary conditions should be fixed once and for
all within the class satisfying the requirement described here above: our
choice is the simplest one satisfying it. Now one can note that, according
to Eqs. (\ref{sessea5.2}) and (\ref{sessea5.2.1}), the integration constant $%
E_{0}$ is fixed in terms of $n$ through the relation 
\begin{equation}
\int_{0}^{n\pi }\left\{ \frac{Y\left( \alpha \right) }{\left[ E_{0}-W(\alpha
)\right] }\right\} ^{1/2}d\alpha =\pm 2\pi \ .  \label{intconstcond}
\end{equation}%
It is easy to see that the above equation for $E_{0}$ always has a real
solution as the integrand interpolates from very small absolute values (when 
$E_{0}$ is very large in absolute value) to very large (when $E_{0}$ is such
that the denominator can have zeros). Hence, one can always find values of $%
E_{0}$ able to satisfy Eq. (\ref{intconstcond}).

\subsection{Baryonic crystals at finite density and its superconducting
nature}

From Eq. \eqref{Tmunu} one can compute the energy density $\mathcal{E}$ of
the configurations defined in Eqs. \eqref{Minkowski}, \eqref{good1} and %
\eqref{good6}, and this turns out to be 
\begin{equation}
\mathcal{E}=4m^{2}\cos {(\alpha )}+\frac{K}{2L^{2}}\tilde{T}_{00}^{\text{Sk}%
}+c_{6}\frac{2q^{2}}{L^{6}}\tilde{T}_{00}^{(6)}+c_{8}\frac{q^{2}}{L^{8}}\sin
^{4}{(\alpha )}\tilde{T}_{00}^{(8)}+\tilde{T}_{00}^{\text{U(1)}}\ ,
\label{EDfull}
\end{equation}%
where 
\begin{align*}
\tilde{T}_{00}^{\text{Sk}}=& \alpha ^{\prime 2}+2\sin ^{2}{(\alpha )}\sin
^{2}(q\theta )(p-2Lu)^{2}+q^{2}\sin ^{2}(\alpha ) \\
& +\frac{\lambda }{L^{2}}\sin ^{2}{(\alpha )}\biggl[(q^{2}+2\sin
^{2}(q\theta )(p-2Lu)^{2})\alpha ^{\prime 2}+2q^{2}\sin ^{2}(q\theta
)(p-2Lu)^{2}\biggl]\ \ , \\
\tilde{T}_{00}^{(6)}=& \sin ^{4}{(\alpha )}\sin ^{2}(q\theta
)(p-2Lu)^{2}\alpha ^{\prime 2}\ ,\quad \tilde{T}_{00}^{(8)}=(\alpha ^{\prime
2}+\sin ^{2}{(\alpha )})(p-2Lu)^{2}\sin ^{2}(q\theta )\alpha ^{\prime 2}-%
\frac{q^{2}}{4}\alpha ^{\prime 4}\ , \\
\tilde{T}_{00}^{\text{U(1)}}& =\frac{1}{L^{2}}\biggl((\partial
_{r}u)^{2}+(\partial _{\theta }u)^{2}\biggl)\ .
\end{align*}%
It is interesting to note that, despite the fact that the term $\mathcal{L}%
_{6}$\ does not contribute to the field equations (as it has been already
emphasized), it does contribute to the energy-momentum tensor. In order to
have a positive definite energy density a necessary condition is $c_{6}\geq 0
$.

On the other hand, the U(1) current in Eq. \eqref{Jmu}, in the ansatz
defined by Eqs. \eqref{Minkowski}, \eqref{good1} and \eqref{good6}, is 
\begin{equation}
J_{\mu }\ =\ \frac{2}{L^{4}}\sin ^{2}{\alpha }\sin ^{2}(q\theta )\Omega
(\alpha )(\partial _{\mu }\Phi -2A_{\mu })\ ,  \label{supercurr1}
\end{equation}%
with $\Omega (\alpha )$ defined in Eq. \eqref{Omegadef}. From the expression
of the current in Eq. \eqref{supercurr1} (see Appendix I for the explicit
form of the components of the current) the following observations are
important.

\textbf{1)} The current does not vanish even when the electromagnetic
potential vanishes ($A_{\mu}=0$).

\begin{figure}[hbtp]
\centering
\includegraphics[scale=.255]{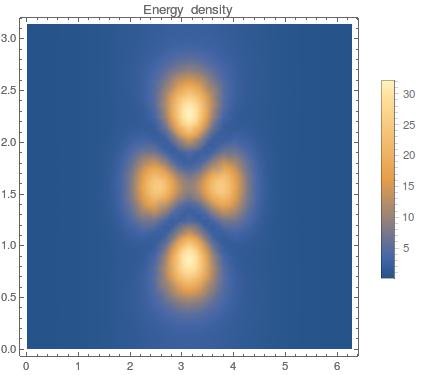}\ \ \includegraphics[scale=.25]{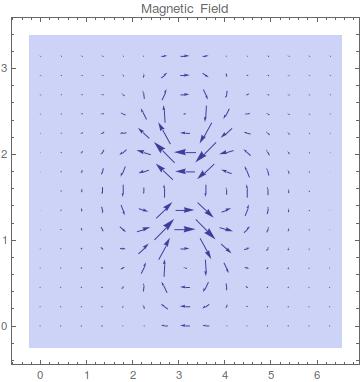} \
\ \includegraphics[scale=.25]{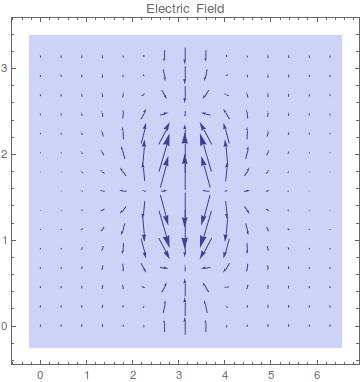} \ \ \includegraphics[scale=.25]{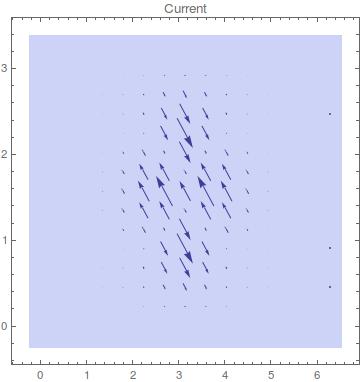}
\caption{From left to right we can see the energy density $\mathcal{E}$, the
magnetic field, the electric field and the current $J_\protect\mu$ for a
configuration with Baryon charge $B=1$. The electric and magnetic fields
vanish at the peaks of the energy density while the current takes its
maximum value.}
\label{Fig1}
\end{figure}

\begin{figure}[hbtp]
\centering
\includegraphics[scale=.255]{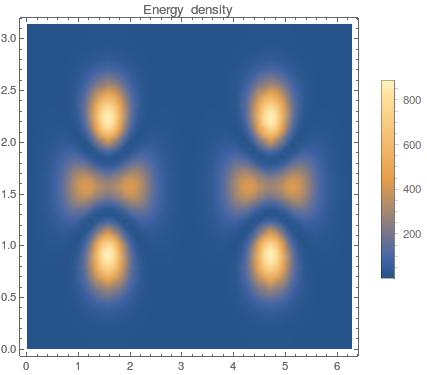}\ \ \includegraphics[scale=.25]{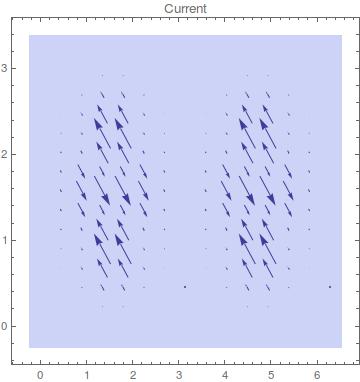}
\ \ \includegraphics[scale=.255]{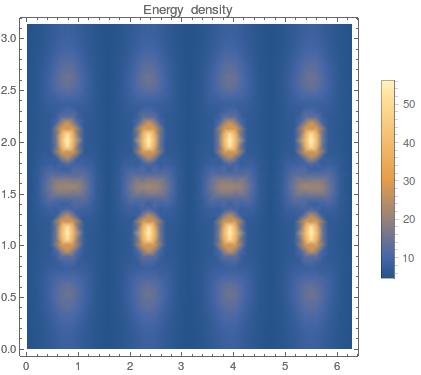} \ \ %
\includegraphics[scale=.25]{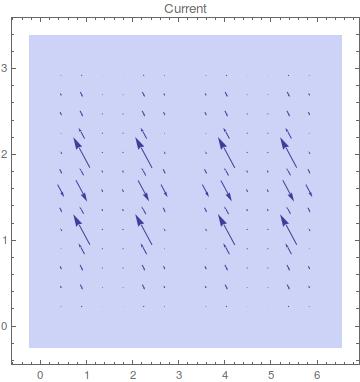}
\caption{From left to right, crystals of superconducting strings with $B=2$
and $B=4$, respectively. The current is concentrated in the tube-shaped
regions defined where (most of) the $\mathcal{E}$ is contained, and vanishes
outside the tubes. The maximum values of $\mathcal{E}$ and the current
coincide in the lattice, which is periodic in $r$, $\protect\theta$ and
perpendicular to the $\protect\phi$ direction, along which the strings
exist. }
\label{Fig2}
\end{figure}

\textbf{2)} Such a \textquotedblleft left over'' 
\begin{equation*}
J_{\mu }^{(0)}=\left. J_{\mu }\right\vert _{A_{\mu }=0}= \frac{2}{L^4}%
\sin^2(\alpha)\sin^2(q\theta)\Omega(\alpha)\partial_\mu \Phi \ ,
\end{equation*}
is maximal where the energy density is maximal and vanishes rapidly far from
the peaks as the plots show (see Fig. \ref{Fig1} and Fig. \ref{Fig2}).

\textbf{3)} $J_{(0)\mu}$ \textit{cannot be turned off continuously}. One can
try to eliminate $J_{(0)\mu}$ either deforming $\alpha$ and/or $\theta$ to
integer multiples of $\pi$ (but this is impossible as such a deformation
would kill the topological charge as well) or deforming $\Phi$ to a constant
(but also this deformation cannot be achieved for the same reason). Note
also that, as it happens in \cite{wittenstrings}, $\Phi$ is defined modulo $%
2\pi$ (as the SU(2) valued field $U$ depends on $\cos\Phi$ and $\sin\Phi$
rather than on $\Phi$\ itself). This implies that $J_{(0)\mu}$ defined in
Eq. (\ref{supercurr1}) is \textit{a superconducting current supported by the
present gauged tubes}. Moreover, these properties are not spoiled by any of
the higher order corrections, parameterized by $c_{p}$.

The plots of the energy density and the current clarify the physical
interpretation of the present multi-solitonic configurations. In Fig. \ref%
{Fig1} and Fig. \ref{Fig2} we have chosen $K=2$, $\lambda=1$, $c_6=c_8=\frac{%
1}{5}$, $m=0$ and $q=p=1$. 
The components of the electric and magnetic fields can be also computed and
are given by 
\begin{gather*}
E_r= -\partial_r u \ , \quad E_\theta= -\partial_\theta u \ , \quad E_\phi=
0 \ , \\
B_r= \frac{1}{L^3}\partial_\theta u \ , \quad B_\theta= -\frac{1}{L^3}%
\partial_r u \ , \quad B_\phi= 0 \ .
\end{gather*}

\subsection{About the existence of exact crystals and the \textit{%
universality} of the ansatz}

In the previous sections we have shown that the low energy limit of QCD
supports the existence of crystals of superconducting Baryonic tubes. Of
course, this result is very technical in nature and, a priori, it is not
clear whether or not one could have expected the appearance (in the low
energy limit of QCD) of topological defects supporting superconducting
currents. Here we will give an intuitive argument which justifies why one
should have expected the existence of such defects.

The first necessary (but, in general, not sufficient) condition that must be
satisfied in order to support the existence of superconductive currents in a
relativistic context is the existence of a massless excitation which can be
coupled consistently to a U(1) gauge field (see the pioneering paper \cite%
{wittenstrings}).

According to Eqs. \eqref{sessea1} and \eqref{sessea2} the SU(2) valued
Skyrme field $U$ describes the dynamical evolution of three scalar degrees
of freedom $\alpha $, $\Theta $ and $\Phi $ which are coupled through the
non-linear kinetic terms typical of Skyrme like models (see Eq. (\ref%
{Iexplicit}) which is the explicit expression of the Skyrme action in terms
of $\alpha $, $\Theta $ and $\Phi $: of course such action is equivalent to
the usual one written implicitly in terms of the SU(2) valued field $U$).
This fact hides a little bit which is the ``best candidate" to carry a
superconductive current since our intuition is built on models where the
interactions appear in potential terms (like in the Higgs model or in the
Ginzburg-Landau model) and not in ``generalized kinetic terms" as in the
present case. So, the question is: how can we decide a priori which whether
or not there is an excitation able to carry a persistent current? In other
words, which one of the three degrees of freedom $\alpha $, $\Theta $ and $%
\Phi $ associated to the SU(2) valued scalar field $U$ can be a carrier of a
superconductive current? In what follows we will detail the intuitive
arguments that lead us to consider $\Phi $ as the most natural choice.

To illustrate our argument, let us first consider the simpler and well known
case of two scalar fields $\Psi ^{i}$, with $i=1,2$, interacting with a
quartic potential in a SO(2) invariant way, 
\begin{equation}  \label{LPsi}
\mathcal{L}=\partial _{\mu }\Psi _{i}\partial ^{\mu }\Psi ^{i}+\lambda (\Psi
_{i}\Psi ^{i}-v^{2})^{2}\ .
\end{equation}%
In order to disclose which degree of freedom is a natural candidate to carry
a chiral current, we can write 
\begin{equation*}
\Psi _{1}=R(x^{\mu })\sin (\chi (x^{\mu }))\ ,\qquad \Psi _{2}=R(x^{\mu
})\cos (\chi (x^{\mu }))\ ,
\end{equation*}%
then, Eq. \eqref{LPsi} becomes 
\begin{equation}
\mathcal{L}=\partial _{\mu }R\partial ^{\mu }R+R^{2}\partial _{\mu }\chi
\partial ^{\mu }\chi +\lambda (R^{2}-v^{2})^{2}\ .  \label{LPsi2}
\end{equation}%
From Eq. \eqref{LPsi2} it is clear that $R$ can not be chiral field because
of the presence of a non-trivial potential term that only depends on $R$ and
generates a natural mass scale in the dynamics of $R$ (there should be no
characteristic mass scale in a superconducting current). On the other hand, $%
\chi $ (which represents a phase and so is defined only modulo $2\pi $)
describes excitations along the valley of the potential, and, consequently,
is a more suitable candidate to carry a superconductive current. Of course,
all of this is well known in the analysis of the Higgs and Goldstone modes,
but this short review helps to identify the correct chiral field in our case
in which there is no potential to look at (as the interactions happen in
non-linear kinetic-like terms). Moreover, the above Lagrangian can also be
naturally coupled to a U(1) gauge field as follows: 
\begin{gather}
\mathcal{L}_{\text{U(1)}} =\partial _{\mu }R\partial ^{\mu }R+R^{2}\left(
\partial _{\mu }\chi -eA_{\mu }\right) \left( \partial ^{\mu }\chi -eA^{\mu
}\right) +\lambda (R^{2}-v^{2})^{2} -F_{\mu \nu }F^{\mu \nu }.
\label{gaugedexample1}
\end{gather}%
It is easy to see that the U(1) current $J_{\mu }$ arising from the above
action is proportional to $J_{\mu }\sim \left( \partial _{\mu }\chi -eA_{\mu
}\right) $. These are part of the main ingredients of \cite{wittenstrings}
to build topological defects supporting superconducting currents. Hence, the
fingerprints to identify the degree of freedom (call it for convenience $%
\Phi ^{\ast }$) suitable to carry the superconducting current are, \textit{%
firstly}, that such a degree of freedom $\Phi ^{\ast }$ should only appear
with a kinetic term in the full action of the theory (as $\chi $ in Eq. (\ref%
{LPsi2}) here above) and without any other explicit non-linear term
involving $\Phi ^{\ast }$ itself\footnote{%
Thus, when one set to constant values all the other degrees of freedom of
the full action (but $\Phi ^{\ast }$ itself) then $\Phi ^{\ast }$ behaves as
a massless field. Indeed, this is the case for the field $\chi $\ in the
above example.}. \textit{Secondly}, the coupling of the theory with a U(1)
gauge theory should only affect the kinetic term of the field $\Phi ^{\ast }$%
. Clearly, the above requirements allow to identify $\chi $ as the field $%
\Phi ^{\ast }$ candidate to be a carrier of a superconducting current in the
above example.

What happens in the present case? Obviously, in the Skyrme case there is no
interaction potential which is responsible for the interactions term: in all
the Skyrme-like models the non-linear behavior is related with generalized
kinetic terms. Nevertheless, as can be seen in Eq. \eqref{Iexplicit}, the $%
\alpha $ and $\Theta $ fields have explicit non-linear interaction terms,
all of them proportional to $\sin ^{2}\alpha $ and/or $\sin ^{2}\Theta $.
Hence (although in the present case there is no potential which clearly
allows to identify the ``proper valleys and Goldstone modes"), it is clear
that neither $\alpha $ nor $\Theta $ can be the analogue of $\chi $ in the
previous example. The reason is that if one sets $\alpha $ to a generic
constant value $\Theta $ will still have non-linear interaction terms and 
\textit{viceversa} if one sets $\Theta $ to a generic constant value $\alpha 
$ will still have non-linear interaction terms\footnote{%
Here, ``generic constant values" mean different from $n\pi $ as otherwise
there would be no kinetic term at all in the action (as it happens when one
sets $R=0$ in the action in Eq. (\ref{LPsi2})). Thus such values are not
relevant for the present analysis.}. Consequently, the fields $\alpha $ and $%
\Theta $ are rather similar to the field $R$ than to the field $\chi $ in
the previous example. The field $\Phi $ on the other hand, has precisely the
same characteristics as the field $\chi $ in the previous example: it is
only defined modulo $2\pi $ (since $U$ depends on $\Phi $ only through $\sin
\Phi $ and $\cos \Phi $) and moreover it appears in the action only with the
corresponding kinetic term. Thus, if you set the other fields $\alpha $ and $%
\Theta $ to generic constant values, then the field $\Phi $ can behave as a
chiral massless field. Thus, a priori, one should have expected that also in
the (generalized) Skyrme model superconducting Baryonic tubes should be
present. Moreover, the minimal coupling of the (generalized) Skyrme model(s)
with the Maxwell theory is defined by the following covariant derivative: $%
\nabla _{\mu }\rightarrow D_{\mu }=\nabla _{\mu }+A_{\mu }[\tau _{3},\cdot ]$%
. From the viewpoint of the $\alpha $, $\Theta $ and $\Phi $ the minimal
coupling rule is completely equivalent to change in the action only the
derivatives of $\Phi $ as follows $\nabla _{\mu }\Phi \rightarrow D_{\mu
}\Phi =\nabla _{\mu }\Phi -2A_{\mu }\Phi $. Hence, also from the viewpoint
of the interaction with the Maxwell theory, the field $\Phi $ is the
analogue of $\chi $ in the previous example and this is exactly what we
need, according to Witten \cite{Witten}, to have a superconducting current.

As a last remark, at a first glance, one could also argue that the presence
of a mass term for the Pions should destroy superconductivity. In fact, this
is not the case since, in terms of $\alpha $, $\Theta $ and $\Phi $, the
mass term for the Pions is $m_{\pi }^{2}(1-\cos \alpha )$, and it only
affects $\alpha $ (in the same way as a mass term in the previous example
would only affect $R$ but would not set a mass scale for $\chi $).

These are the intuitive arguments which strongly suggest a priori that it
certainly pay off to look for superconducting solitons in the generalized
Skyrme model(s) and that $\Phi $ should be the superconducting carrier.

Furthermore, by requiring that $\nabla _{\mu }\Phi \nabla ^{\mu }\Phi $
vanishes (as it is expected for chiral fields), the field equations are
enormously simplified (see Eqs. \eqref{equ1}, \eqref{equ2} and \eqref{equ3}
on Appendix II). This simplification occurs not only on the Skyrme model
case, but even if higher derivative order terms are considered, as we have
already discussed.

\subsection{Stability analysis}

One of the most intriguing results of the present framework is that the
physical properties of these superconducting Baryonic tubes remain the same
no matter how many subleading terms are included in the generalized Skyrme
model\footnote{%
This construction also works when further corrections from chiral
perturbation theory are considered \cite{Zahed2}. Indeed, even if one
includes the quartic corrections considered in \cite{Zahed2} (in which the
anti-commutators between the Maurer-Cartan forms appear), the SU(2) field
equations still reduce to a single integrable first-order ODE. We will not
discuss these terms explicitly since they do not change the qualitative
picture presented here.}. In other words, these topologically non-trivial
configurations are almost \textquotedblleft \textit{theory independent}". As
it has been already emphasized, this happens since the ansatz defined in Eq.
(\ref{good1}) \textit{works in exactly the same way without any change at all%
} no matter how many higher order terms are included in the generalized
Skyrme action. In particular, the field equations will always be reduced to
a single integrable ODE for $\alpha $ and the corresponding configurations
will describe superconducting tubes. Hence, the present topological gauged
solitons are likely to be a universal feature of QCD \textit{as they stay
the same at any order in the large} \textbf{N} \textit{expansion}.

\begin{figure}[ht]
\centering
\includegraphics[scale=.22]{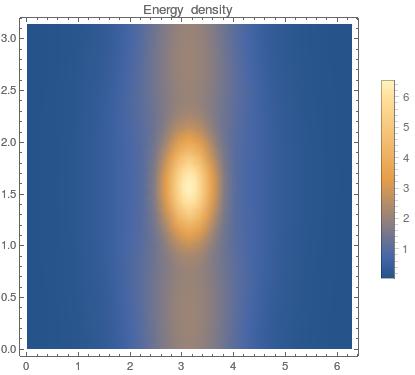}\ %
\includegraphics[scale=.22]{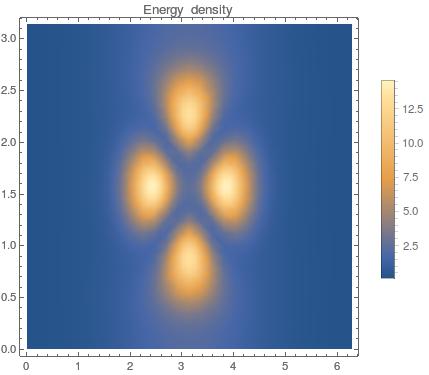}\ %
\includegraphics[scale=.22]{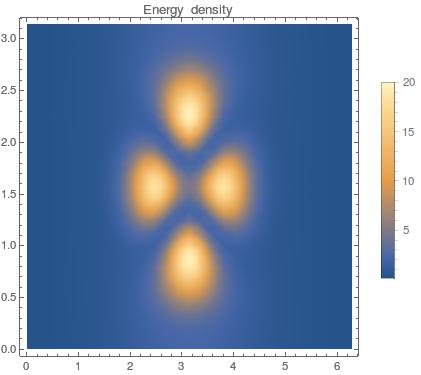}\ \includegraphics[scale=.22]{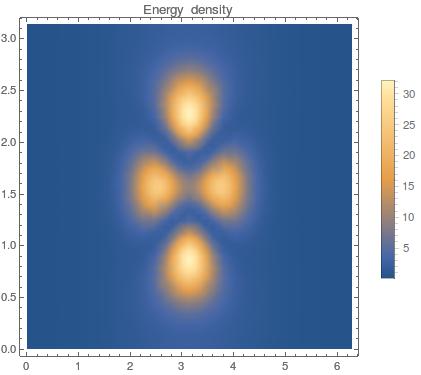}
\ \includegraphics[scale=.22]{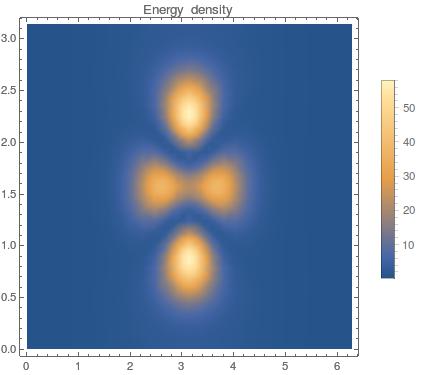}
\caption{Energy density for different configurations with $B=1$. From left
to right: energy density only including the non-linear sigma model
contribution, then up to the Skyrme contribution, up to $\mathcal{L}_6$, up
to $\mathcal{L}_8$ and finally up to order $\mathcal{L}_{12}$ .}
\label{Fig3}
\end{figure}

\begin{figure}[ht]
\centering
\includegraphics[scale=.285]{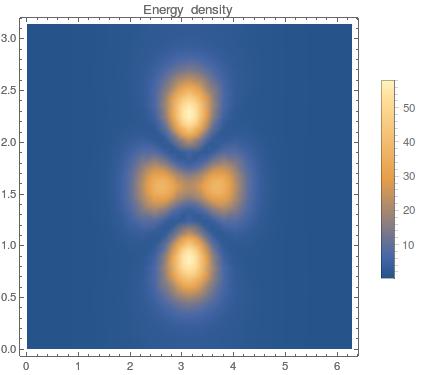}\ \includegraphics[scale=.28]{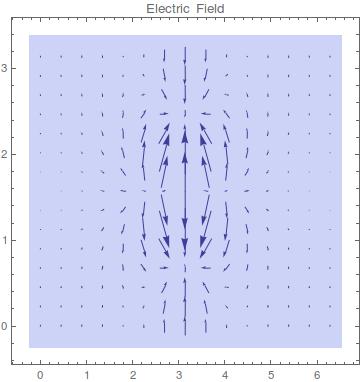}\ %
\includegraphics[scale=.28]{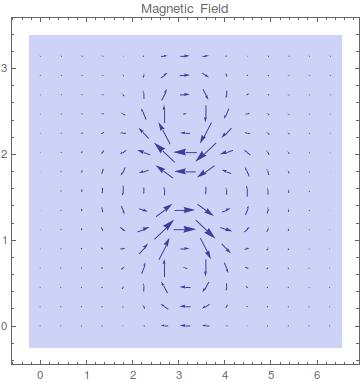}\newline
\includegraphics[scale=.28]{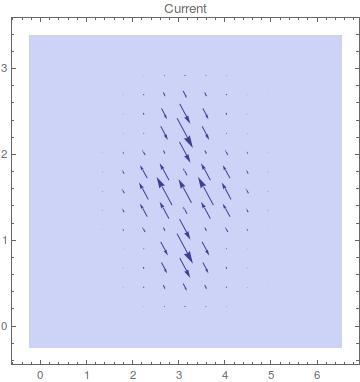}\ \includegraphics[scale=.285]{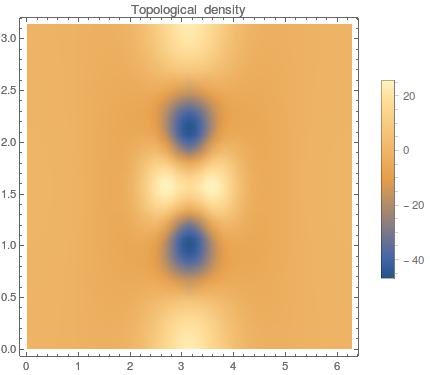}
\ \ \includegraphics[scale=.34]{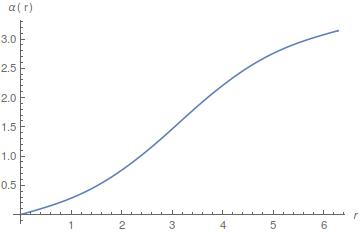}
\caption{From left to right: Energy density $\mathcal{E}$, electric field,
magnetic field, current, topological density $\protect\rho_\text{B}$ and
radial profile $\protect\alpha(r)$, up to the contribution $\mathcal{L}_{12}$
for the configuration with $B=1$.}
\label{Fig4}
\end{figure}

To give a flavor of why such a property is so surprising, let us consider
the 't Hooft-Polyakov monopole \cite{monopole1}, \cite{monopole2}. The
stability of these configurations in the Georgi-Glashow model is of course
very well understood. However, if one deforms even slightly the theory, the
properties of the 't Hooft-Polyakov monopole are going to change as well
(see, for instance, \cite{monopole3} and references therein). To give just
an example: in \cite{monopole3} the authors considered a very natural
correction to the Georgi-Glashow model which leads to a non-spherical
deformation of the 't Hooft-Polyakov monopole (so that, in particular, the
ansatz for the 't Hooft-Polyakov monopole must be changed accordingly).
Consequently, the shape of non-Abelian monopoles is also going to change
when these types of deformations of the Yang-Mills theory are included. On
the other hand, the superconducting Baryonic tubes constructed here keep
their properties at any order in the large \textbf{N} expansion. To the best
of authors knowledge, these are the first examples of ``universal" gauged
solitons in the low energy limit of QCD described by an ansatz able to
survive to all the subleading large \textbf{N} corrections. Indeed, the
subleading corrections to the generalized Skyrme model will only change
slightly the plot of $\alpha (r)$ keeping unchanged the plots and the
properties of the superconducting currents and of the energy density (see
Figure \ref{Fig3}). Here below we write the field equation for $\alpha (r)$
with corrections up to order $\mathcal{L}_{12}$ together with the plots of
the energy density of the superconducting tubes in the sector with Baryonic
charge $B=1$ in Figure \ref{Fig4}. For this we have chosen $K = 2$, $\lambda
= 1$, $c_6 = c_8 = c_{12}=\frac{1}{5}$, $m = 0$ and $q = p = 1$.

The field equations are given by 
\begin{gather}
\alpha ^{\prime \prime }-\frac{q^{2}}{2}\sin (2\alpha )+\frac{4m^{2}L^{2}}{K}%
\sin {\alpha }+\frac{\lambda q^{2}}{L^{2}}\sin (\alpha )[\cos (\alpha
)\alpha ^{\prime 2}+\sin (\alpha )\alpha ^{\prime \prime }]  \notag \\
-\frac{3c_{8}q^{4}}{KL^{6}}\sin ^{3}(\alpha )[\cos (\alpha )\alpha ^{\prime
2}+\sin (\alpha )\alpha ^{\prime \prime }]\alpha ^{\prime 2}-\frac{%
15c_{12}q^{6}}{KL^{6}}\sin ^{5}(\alpha )[\cos (\alpha )\alpha ^{\prime
2}+\sin (\alpha )\alpha ^{\prime \prime }]\alpha ^{\prime 4}\ =\ 0\ ,
\label{eqconL12}
\end{gather}%
that can be written again as a first order equation 
\begin{equation}
\left\{ X\left( \alpha \right) \left( \alpha ^{\prime }\right) ^{2}+Z\left(
\alpha \right) +\tilde{E}_{0}\right\} ^{\prime }=0\ ,  \label{sg1}
\end{equation}%
where in this case 
\begin{gather*}
X\left( \alpha \right) =1+\frac{\lambda q^{2}}{L^{2}}\sin ^{2}(\alpha )-%
\frac{3c_{8}q^{4}}{2KL^{6}}\sin ^{4}(\alpha )\alpha ^{\prime 2}-\frac{%
5c_{12}q^{6}}{KL^{10}}\sin ^{6}\alpha (\alpha ^{\prime })^{4}\ , \\
Z(\alpha )=\frac{q^{2}}{2}\cos (2\alpha )-\frac{8m^{2}L^{2}}{K}\cos (\alpha
)\ .
\end{gather*}%
Note also that Eq. (\ref{sg1}) can be seen as a cubic polynomial in the
variable $z=\alpha ^{\prime 2}$ which allows, once again, to reduce the
complete field equations to a simple quadrature of the form:%
\begin{equation*}
\frac{d\alpha }{\widetilde{\chi }^{1/2}}=\pm dr \ ,\qquad \widetilde{\chi }=%
\widetilde{\chi }(\alpha ,E_{0}) \ ,
\end{equation*}%
where $\widetilde{\chi }(\alpha ,E_{0})$ is the positive real root of the
cubic polynomial in $z=\alpha ^{\prime 2}$ defined in Eq. (\ref{sg1}). The
integration constant $\tilde{E}_{0}$ always allows such polynomial to have
positive real roots. 

\subsubsection{Perturbations on the profile}

A remark on the stability of the above superconductive tubes is in order. In
many situations, when the hedgehog property holds (so that the field
equations reduce to a single equation for the profile) the most dangerous
perturbations\footnote{%
``Dangerous perturbations" in the sense that are the perturbations which, in
the most common situations, are more likely to have some negative
eigenvalues.} are perturbations of the profile which keep the structure of
the hedgehog ansatz (see \cite{shifman1}, \cite{shifman2} and references
therein). In the present case these are perturbations of the following type:%
\begin{equation}
\alpha \rightarrow \alpha +\varepsilon \xi \left( r\right) \ ,\ \ \
0<\varepsilon \ll 1\ ,  \label{pert}
\end{equation}%
which do not change the Isospin degrees of freedom associated with the
functions $\Theta $\ and $\Phi $. A direct computation reveals that the
linearized version of Eq. (\ref{sessea4}) around a background solution $%
\alpha _{0}\left( r\right) $ of Baryonic charge $B=np$ always has the
following zero-mode: $\xi \left( r\right) =\nabla_{r}\alpha _{0}\left(
r\right) $. Due to the fact that the integration constant $E_{0}$ (defined
in Eqs. (\ref{sessea5}), (\ref{sessea5.2}) and (\ref{sessea5.2.1})) can
always be chosen in such a way that $\nabla_{r}\alpha _{0}\left( r\right) $
never vanishes, the zero mode $\xi \left( r\right) $ has no node so that it
must be the perturbation with lowest energy. Thus, the present solutions are
stable under the above potentially dangerous perturbations. Although this is
not a complete proof of stability, it is a very non-trivial test.

\subsubsection{Electromagnetic perturbations}

A very useful approach to study the stability of the superconducting
Baryonic tubes is to perform electromagnetic perturbations on the effective
medium defined by the topological solitons. This is a good approach in the
't Hooft limit, since in the semiclassical interaction Photon-Baryon, the
Baryon is essentially unaffected due to the Photon has zero mass (see \cite%
{WeigelNotes}).

The complete stability analysis requires to study the most general
perturbations of the solutions defined in Eqs. (\ref{good1}), (\ref%
{sessea5.2}) and (\ref{sessea5.2.1}). This is a very hard task even
numerically as it involves a coupled system of linear PDEs, therefore in
practical terms, consider only electromagnetic perturbations greatly
simplifies the stability analysis and allows to reveal very relevant
features of the superconducting tubes, as we will see immediately.

Here we will analyze the simplest non-trivial case which is related to the
role of the subleading corrections in the 't Hooft expansion to the Skyrme
model of the sixth order. As it has been already emphasized in the previous
sections such sixth order term does not even appear in the equation for the
profile (while it does enter in the corresponding Maxwell equations). This
is very interesting since it shows that, despite the universal character of
the present crystals of gauged solitons (which are almost unaffected by the
subleading terms), their stability properties may depend explicitly on the
subleading terms themselves. Also for simplicity reasons, we will set $m$ to
zero.

Let us consider the following perturbations on the Maxwell potential 
\begin{equation*}
(u,0,0,-Lu)\rightarrow (u+\varepsilon \xi _{1},0,0,-Lu+\varepsilon \xi
_{2})\ ,\qquad \xi _{i}=\xi _{i}(t,r,\theta ,\phi )\ ,\qquad 0<\varepsilon
\ll 1\ .
\end{equation*}%
At first order in the parameter $\varepsilon $ the Maxwell equations become 
\begin{gather*}
\partial _{\theta }(\partial _{\phi }\xi _{2}-L^{2}\partial _{t}\xi _{1})=0\
, \\
\partial _{r}(\partial _{\phi }\xi _{2}-L^{2}\partial _{t}\xi _{1})=0\ , \\
(\partial _{r}^{2}+\partial _{\theta }^{2}+\partial _{\phi }^{2})\xi
_{1}-\partial _{\phi }\partial _{t}\xi _{2}+V\xi _{1}\ =\ 0\ , \\
(\partial _{r}^{2}+\partial _{\theta }^{2}-L^{2}\partial _{t}^{2})\xi
_{2}+L^{2}\partial _{\phi }\partial _{t}\xi _{1}+V\xi _{2}\ =\ 0\ ,
\end{gather*}%
where $V$ and $\Omega (\alpha )$ are defined in Eqs. \eqref{Omegadef} and %
\eqref{Vdef} (up to sixth order). Note that, since we want to test linear
stability one should check the (absence of) growing modes in time. This
implies that $\xi _{1}$ and $\xi _{2}$ must depend on the temporal
coordinate. But, according to the previous equations if $\xi _{1}$ and $\xi
_{2}$ depend on time these functions must also depend on the coordinate $%
\phi $, that is 
\begin{equation*}
\partial _{t}\xi _{i}\neq 0\ ,\qquad \partial _{\phi }\xi _{i}\neq 0\
,\qquad \xi _{i}=\{\xi _{1},\xi _{2}\}\ .
\end{equation*}%
We can assume that 
\begin{equation*}
\partial _{\phi }\xi _{2}=L^{2}\partial _{t}\xi _{1}\ .
\end{equation*}%
By consider the Fourier transformation in the coordinate $\phi $ we get an
equation for $\widehat{\xi }_{i}$ in the form 
\begin{equation*}
-\Box \widehat{\xi }_{i}+(k^{2}-V)\widehat{\xi }_{i}=0\ ,\qquad \Box \equiv
-L^{2}\partial _{t}^{2}+\partial _{r}^{2}+\partial _{\theta }^{2}\ ,
\end{equation*}%
where 
\begin{equation*}
\widehat{\xi }_{i}(t,r,\theta ,k)=\int \xi _{i}(t,r,\theta ,\phi )e^{-ik\phi
}d\phi \ ,
\end{equation*}%
is the Fourier transform of $c_{i}$, and the non-vanishing eigenvalue $%
k=l/(2\pi )$ is the wave-number along the $\phi $-direction, with $l$ a
non-vanishing integer.

According to Duhamel's principle (see, for instance, \cite{Sternberg} and
references therein), an inhomogeneous equation for a function $W=W(x,t)$ of
the form 
\begin{equation*}
(d_{t}^{2}+M)W=f\ ,
\end{equation*}%
with $M$ a non-negative operator and initial conditions $W(\cdot ,0)=\psi
_{1}$, $\partial _{t}W(\cdot ,0)=\psi _{2}$, has the following general
solution 
\begin{equation*}
W(\cdot ,t)=\partial _{t}B(t)\psi _{1}+B(t)\psi _{2}+\int_{0}^{t}B(t-\tau
)f(\tau )d\tau \ ,\quad B(t)=M^{-\frac{1}{2}}\sin (tM^{\frac{1}{2}})\ .
\end{equation*}%
In our case, to ensure that the perturbed Maxwell equation can be solved we
need to demand that $V_{\text{eff}}>0$, with 
\begin{equation}
V_{\text{eff}}=k^{2}-V\ .  \label{effpot1}
\end{equation}%
Since $V$ depends on $\alpha $ and the square of its derivative $\alpha
^{\prime }=\alpha ^{\prime }(E_{0})$, defined via Eq. \eqref{sessea5}, one
can find the following upper bound to the potential $V$: 
\begin{equation*}
\left\vert V\right\vert \leq V_{\text{max}}<\frac{4}{L^{2}}\biggl(%
KL^{4}+(q^{2}+E_{0})\lambda KL^{2}+2q^{2}c_{6}E_{0}\biggl)\ .
\end{equation*}%
Then, a necessary condition\footnote{%
We think that this bound can be improved increasing the stability range of
these solutions. Since, in the present subsection, we only want to show that
the subleading terms could have relevant physical effects, we will not
discuss the improved bound for the effective potential further. We hope to
come back on this issue in a future publication.} for a positive defined
effective potential $V_{\text{eff}}$ in Eq. (\ref{effpot1}) is%
\begin{equation}
\left( \frac{l}{2\pi }\right) ^{2}-\frac{4}{x}\biggl(Kx^{2}+(q^{2}+E_{0})%
\lambda Kx+2q^{2}c_{6}E_{0}\biggl)\geq 0\ ,  \label{effpot2}
\end{equation}%
\begin{equation*}
x=L^{2}\ ,
\end{equation*}%
(the most restrictive case being the one with $l^{2}=1$ as it is easier to
satisfy the above inequality when $l^{2}$ is large). The above inequality
set an upper bound on the allowed values of $L$ (which is the same as a
lower bound on the allowed values of Baryon densities):%
\begin{equation}
\frac{x}{16\pi ^{2}}-\biggl(Kx^{2}+(q^{2}+E_{0})\lambda Kx+2q^{2}c_{6}E_{0}%
\biggl)\geq 0\ .  \label{effpot3}
\end{equation}%
The above inequality is equivalent to%
\begin{equation}
L_{\text{Min}}^{2}\leq x\leq L_{\text{Max}}^{2}\ ,  \label{effpot4}
\end{equation}%
\begin{eqnarray}
L_{\text{Max}}^{2} &=&\frac{1}{2K}\biggl(\frac{1}{16\pi ^{2}}%
-(E_{0}+q^{2})K\lambda +\sqrt{\biggl(\frac{1}{16\pi ^{2}}-(E_{0}+q^{2})K%
\lambda \biggl)^{2}-8c_{6}q^{2}E_{0}K}\biggl)\ ,  \label{effpot4.1} \\
L_{\text{Min}}^{2} &=&\frac{1}{2K}\biggl(\frac{1}{16\pi ^{2}}%
-(E_{0}+q^{2})K\lambda -\sqrt{\biggl(\frac{1}{16\pi ^{2}}-(E_{0}+q^{2})K%
\lambda \biggl)^{2}-8c_{6}q^{2}E_{0}K}\biggl)\ ,  \label{effpot4.2}
\end{eqnarray}%
together with the obvious condition that%
\begin{equation}
x\geq 0\ .  \label{effpot5}
\end{equation}%
Thus, in the range of parameters in which $L_{\text{Max}}^{2}$ is positive
(which always exists) the conditions on $L^{2}$ is 
\begin{equation*}
L^{2}<\frac{1}{2K}\biggl(\frac{1}{16\pi ^{2}}-(E_{0}+q^{2})K\lambda +\sqrt{%
\biggl(\frac{1}{16\pi ^{2}}-(E_{0}+q^{2})K\lambda \biggl)%
^{2}-8c_{6}q^{2}E_{0}K}\biggl)\ .
\end{equation*}%
Thus, at a first glance, from Eq. \eqref{sessea4} one could think that the
presence of the $c_{6}$, which for energetic considerations must be positive
(see Eq. \eqref{EDfull}), do not play any role in the perturbation of the
system. However, it is quite interesting to see that this term can in fact
affect the stability of the system determining the maximum allowed value of
the size of the box in which the superconducting strings are confined. We
hope to come back on the physical properties of these gauged crystals in the
low energy limit of QCD in a future publication.

\section{Conclusions and perspectives}

The Maxwell-gauged Skyrme model in $(3+1)$-dimensions together with all the
subleading corrections in the 't Hooft expansion admit configurations
describing ordered arrays of Baryonic superconducting tubes where (most of)
the Baryonic charge and total energy is concentrated in tube-shaped regions.
The corresponding current cannot be deformed continuously to zero as it is
tied to the topological charge. Quite remarkably, no matter how many
subleading terms are included, these ordered arrays of gauged solitons are
described by the very same ansatz and keep their main properties manifesting
a sort of universal character. The similarity with the plots obtained
numerically in the analysis of nuclear spaghetti phase is quite amusing \cite%
{pasta1}. These results open the unprecedented possibility to analyze these
complex structures with analytic tools which are able to disclose novel
features which are difficult to analyze with many body simulations. On the
other hand, the subleading terms in the 't Hooft expansion (which almost do
not affect the solutions of the field equations) do, in fact, affect the
stability properties of the superconducting tubes setting upper bounds on
the allowed values of the spatial volume in which they can live.

\subsection*{Acknowledgments}

F. C. has been funded by Fondecyt Grants 1200022. M. L. and A. V. are funded
by FONDECYT post-doctoral grant 3190873 and 3200884. The Centro de Estudios
Cient\'{\i}ficos (CECs) is funded by the Chilean Government through the
Centers of Excellence Base Financing Program of Conicyt.

\section*{Appendix I: Explicit tensors}

\label{AppI}

From the ansatz defined in Eqs. \eqref{Minkowski}, \eqref{sessea1}, %
\eqref{sessea2}, \eqref{good1} and \eqref{good6}, the explicit expression of
the matrix $U$, the components of the tensor $L_{\mu}$ and the components of
the current $J_{\mu }$ are given by 
\begin{equation*}
U=%
\begin{pmatrix}
\cos(\alpha)+i\sin(\alpha)\cos(q\theta) & i e^{-ip(t/L-\phi)} \sin(\alpha)
\sin(q\theta) \\ 
i e^{ip(t/L-\phi)} \sin(\alpha) \sin(q\theta) & \cos(\alpha)-i\sin(\alpha)%
\cos(q\theta)%
\end{pmatrix}%
\ ,
\end{equation*}
\begin{equation*}
L_{t}=(\frac{p}{L}-2 u)%
\begin{pmatrix}
i\sin ^{2}\alpha \sin ^{2}(q\theta ) & e^{-ip(t/L-\phi)}\sin \alpha \sin
(q\theta )(\cos \alpha -i\sin \alpha \cos (q\theta ) \\ 
-e^{ip(t/L-\phi)}\sin \alpha \sin (q\theta )(\cos \alpha +i\sin \alpha \cos
(q\theta ) & -i\sin ^{2}\alpha \sin ^{2}(q\theta )%
\end{pmatrix}%
\ ,
\end{equation*}%
\begin{equation*}
L_{r}=%
\begin{pmatrix}
i\cos (q\theta )\alpha ^{\prime } & e^{-ip(t/L-\phi)}\sin (q\theta )\alpha
^{\prime } \\ 
ie^{ip(t/L-\phi)}\sin (q\theta )\alpha ^{\prime } & -i\cos (q\theta )\alpha
^{\prime }%
\end{pmatrix}%
\ ,
\end{equation*}%
\begin{equation*}
L_{\theta }=%
\begin{pmatrix}
-iq\sin \alpha \cos \alpha \sin (q\theta ) & qe^{-ip(t/L-\phi)}\sin \alpha
(\sin \alpha +i\cos \alpha \cos (q\theta )) \\ 
-qe^{ip(t/L-\phi)}\sin \alpha (\sin \alpha -i\cos \alpha \cos (q\theta )) & 
iq\sin \alpha \cos \alpha \sin (q\theta )%
\end{pmatrix}%
\ ,
\end{equation*}%
\begin{equation*}
L_{\phi}=-L L_{t}\ ,
\end{equation*}
\begin{gather*}
J_{t}= \frac{2}{L^4}\sin^2(\alpha)\sin^2(q\theta)\biggl( KL^{2}(L^{2}+q^{2}%
\lambda \sin ^{2}(\alpha )+\lambda \alpha ^{\prime 2})+q^{2}\sin ^{2}(\alpha
)\alpha ^{\prime 2}\biggl[2c_{6}+\frac{c_{8}}{L^2}(q^2\sin^2(\alpha)+%
\alpha^{\prime 2}) \biggl] \biggl) \\
\times \biggl(\frac{p}{L}-2u\biggl) \ , \\
J_{r}=J_{\theta }=0 \ , \quad J_\phi=-L J_t \ .
\end{gather*}

\section*{Appendix II: Reducing the Skyrme equations}

\label{AppII}

In this appendix we will show how and why the Skyrme equations (using the
ansatz defined by Eqs. \eqref{Minkowski}, \eqref{sessea1}, \eqref{sessea2},
and \eqref{good1}) are reduced to just one integrable ODE for the soliton
profile $\alpha =\alpha (r)$. To see this fact it is possible to take two
paths, as we detail below.

In order to make this reduction clear, in this appendix we will consider the
action in Eq. \eqref{sky1} without the higher order terms and without the
coupling with Maxwell theory, i.e. we will deal only with the usual Skyrme
action 
\begin{gather}
I_{\text{Skyrme}}=\int \frac{d^{4}v}{4}\left[ K\mathrm{Tr}\left( L^{\mu
}L_{\mu }+\frac{\lambda }{8}G_{\mu \nu }G^{\mu \nu }\right) \right] \ ,
\label{ISk} \\
L_{\mu }=U^{-1}\nabla _{\mu }U\ ,\quad G_{\mu \nu }=[L_{\mu },L_{\nu }]\
,\quad d^{4}v=d^{4}x\sqrt{-g}\ ,  \notag \\
U\in \text{SU(2)}\ ,\ \ L_{\mu }=L_{\mu }^{j}t_{j}\ ,\ \ t_{j}=i\sigma _{j}\
.  \notag
\end{gather}%
The reason for doing this is that the mechanism which makes the present
strategy successful with the usual Skyrme model works in exactly the same
way when the higher order terms are included.

The most direct way to see that the Skyrme equations are reduced to just one
equation with the ansatz defined by Eqs. \eqref{Minkowski}, \eqref{sessea1}, %
\eqref{sessea2}, and \eqref{good1} corresponds to write the action in Eq. %
\eqref{ISk} explicitly in terms of the functions $\alpha =\alpha (x^{\mu })$%
, $\Theta =\Theta (x^{\mu })$ and $\Phi =\Phi (x^{\mu })$, according to Eqs. %
\eqref{sessea1}, \eqref{sessea2}. In this parameterization Eq. \eqref{ISk}
becomes 
\begin{equation}
I_{\text{Skyrme}}=\frac{K}{2}\int d^{4}v\left( 
\begin{array}{c}
\nabla _{\mu }\alpha \nabla ^{\mu }\alpha +\sin ^{2}{\alpha }\nabla _{\mu
}\Theta \nabla ^{\mu }\Theta +\sin ^{2}{\alpha }\sin ^{2}{\Theta }\nabla
_{\mu }\Phi \nabla ^{\mu }\Phi \\ 
+\lambda \left( 
\begin{array}{c}
\sin ^{2}{\alpha }\left( (\nabla _{\mu }\alpha \nabla ^{\mu }\alpha )(\nabla
_{\nu }\Theta \nabla ^{\nu }\Theta )-(\nabla _{\mu }\alpha \nabla ^{\mu
}\Theta )^{2}\right) \\ 
+\sin ^{2}{\alpha }\sin ^{2}{\Theta }\left( (\nabla _{\mu }\alpha \nabla
^{\mu }\alpha )(\nabla _{\nu }\Phi \nabla ^{\nu }\Phi )-(\nabla _{\mu
}\alpha \nabla ^{\mu }\Phi )^{2}\right) \\ 
+\sin ^{4}{\alpha }\sin ^{2}{\Theta }\left( (\nabla _{\mu }\Theta \nabla
^{\mu }\Theta )(\nabla _{\nu }\Phi \nabla ^{\nu }\Phi )-(\nabla _{\mu
}\Theta \nabla ^{\mu }\Phi )^{2}\right)%
\end{array}%
\right)%
\end{array}%
\right) \ .  \label{Iexplicit}
\end{equation}%
Now, varying the action in Eq. \eqref{Iexplicit} w.r.t the functions $\alpha 
$, $\Theta $, $\Phi $, in a long but direct calculation, we get to the
following set of equations: 
\begin{equation}
\begin{array}{c}
\left( -\Box \alpha +\sin (\alpha )\cos (\alpha )\left( \nabla _{\mu }\Theta
\nabla ^{\mu }\Theta +\sin ^{2}\Theta \nabla _{\mu }\Phi \nabla ^{\mu }\Phi
\right) \right) \\ 
+\lambda \left( 
\begin{array}{c}
\sin (\alpha )\cos (\alpha )\left( (\nabla _{\mu }\alpha \nabla ^{\mu
}\alpha )(\nabla _{\nu }\Theta \nabla ^{\nu }\Theta )-(\nabla _{\mu }\alpha
\nabla ^{\mu }\Theta )^{2}\right) \\ 
+\sin (\alpha )\cos (\alpha )\sin ^{2}(\Theta )\left( (\nabla _{\mu }\alpha
\nabla ^{\mu }\alpha )(\nabla _{\nu }\Phi \nabla ^{\nu }\Phi )-(\nabla _{\mu
}\alpha \nabla ^{\mu }\Phi )^{2}\right) \\ 
+2\sin ^{3}(\alpha )\cos (\alpha )\sin ^{2}(\Theta )\left( (\nabla _{\mu
}\Theta \nabla ^{\mu }\Theta )(\nabla _{\nu }\Phi \nabla ^{\nu }\Phi
)-(\nabla _{\mu }\Theta \nabla ^{\mu }\Phi )^{2}\right) \\ 
-\nabla _{\mu }\left( \sin ^{2}(\alpha )(\nabla _{\nu }\Theta \nabla ^{\nu
}\Theta )\nabla ^{\mu }\alpha \right) +\nabla _{\mu }\left( \sin ^{2}(\alpha
)(\nabla _{\nu }\alpha \nabla ^{\nu }\Theta )\nabla ^{\mu }\Theta \right) \\ 
-\nabla _{\mu }\left( \sin ^{2}(\alpha )\sin ^{2}(\Theta )(\nabla _{\nu
}\Phi \nabla ^{\nu }\Phi )\nabla ^{\mu }\alpha \right) +\nabla _{\mu }\left(
\sin ^{2}(\alpha )\sin ^{2}(\Theta )(\nabla _{\nu }\alpha \nabla ^{\nu }\Phi
)\nabla ^{\mu }\Phi \right)%
\end{array}%
\right)%
\end{array}%
=0\ ,  \label{equ1}
\end{equation}%
\begin{equation}
\begin{array}{c}
\left( -\sin ^{2}(\alpha )\Box \Theta -2\sin (\alpha )\cos (\alpha )\nabla
_{\mu }\alpha \nabla ^{\mu }\Theta +\sin ^{2}(\alpha )\sin (\Theta )\cos
(\Theta )\nabla _{\mu }\Phi \nabla ^{\mu }\Phi \right) \\ 
+\lambda \left( 
\begin{array}{c}
\sin ^{2}(\alpha )\sin (\Theta )\cos (\Theta )\left( (\nabla _{\mu }\alpha
\nabla ^{\mu }\alpha )(\nabla _{\nu }\Phi \nabla ^{\nu }\Phi )-(\nabla _{\mu
}\alpha \nabla ^{\mu }\Phi )^{2}\right) \\ 
+\sin ^{4}(\alpha )\sin (\Theta )\cos (\Theta )\left( (\nabla _{\mu }\Theta
\nabla ^{\mu }\Theta )(\nabla _{\nu }\Phi \nabla ^{\nu }\Phi )-(\nabla _{\mu
}\Theta \nabla ^{\mu }\Phi )^{2}\right) \\ 
-\nabla _{\mu }\left( \sin ^{2}(\alpha )(\nabla _{\nu }\alpha \nabla ^{\nu
}\alpha )\nabla ^{\mu }\Theta \right) +\nabla _{\mu }\left( \sin ^{2}(\alpha
)(\nabla _{\nu }\alpha \nabla ^{\nu }\Theta )\nabla ^{\mu }\alpha \right) \\ 
-\nabla _{\mu }\left( \sin ^{4}(\alpha )\sin ^{2}(\Theta )(\nabla _{\nu
}\Phi \nabla ^{\nu }\Phi )\nabla ^{\mu }\Theta \right) +\nabla _{\mu }\left(
\sin ^{4}(\alpha )\sin ^{2}(\Theta )(\nabla _{\nu }\Theta \nabla ^{\nu }\Phi
)\nabla ^{\mu }\Phi \right)%
\end{array}%
\right)%
\end{array}%
=0\ ,  \label{equ2}
\end{equation}%
\begin{equation}
\begin{array}{c}
\left( -\sin ^{2}(\alpha )\sin ^{2}(\Theta )\Box \Phi -2\sin (\alpha )\cos
(\alpha )\sin ^{2}(\Theta )\nabla _{\mu }\alpha \nabla ^{\mu }\Phi -2\sin
^{2}(\alpha )\sin (\Theta )\cos (\Theta )\nabla _{\mu }\Theta \nabla ^{\mu }{%
\Phi }\right) \\ 
+\lambda \left( 
\begin{array}{c}
-\nabla _{\mu }\left[ \sin ^{2}(\alpha )\sin ^{2}(\Theta )(\nabla _{\nu
}\alpha \nabla ^{\nu }\alpha )\nabla ^{\mu }\Phi \right] +\nabla _{\mu }%
\left[ \sin ^{2}(\alpha )\sin ^{2}(\Theta )(\nabla _{\nu }\alpha \nabla
^{\nu }\Phi )\nabla ^{\mu }\alpha \right] \\ 
-\nabla _{\mu }\left[ \sin ^{4}(\alpha )\sin ^{2}(\Theta )(\nabla _{\nu
}\Theta \nabla ^{\nu }\Theta )\nabla ^{\mu }\Phi \right] +\nabla _{\mu }%
\left[ \sin ^{4}(\alpha )\sin ^{2}(\Theta )(\nabla _{\nu }\Theta \nabla
^{\nu }\Phi )\nabla ^{\mu }\Theta \right]%
\end{array}%
\right)%
\end{array}%
=0\ .  \label{equ3}
\end{equation}%
The equations system written above are completely equivalent to the system
in Eq. \eqref{sessea0} when the parameterization in Eqs. \eqref{sessea1} and %
\eqref{sessea2} is considered. For instance, one can check that with this
parametrization the original spherical hedgehog ansatz of Skyrme himself 
\cite{skyrme} reads%
\begin{gather}
\alpha =\alpha (x) \ , \quad \Theta =\theta \ ,\quad \Phi =\phi \ ,
\label{ori1} \\
ds^{2} = -dt^{2}+dx^{2}+x^{2}\left( d\theta ^{2}+\sin ^{2}\theta d\phi
^{2}\right) \ ,  \label{ori2}
\end{gather}%
where $x$ is the radial coordinate of flat space-time metric in spherical
coordinates. If one plugs the ansatz in Eqs. (\ref{ori1}) and (\ref{ori2})
into the field equations (\ref{equ1}), (\ref{equ2}) and (\ref{equ3}) one can
see that, first of all, the field equations for $\Theta $ and $\Phi $ are
identically satisfied and, secondly, that the remaining field equation for $%
\alpha $ reduces to the well known equation for the profile of the spherical
Skyrmion. This is the defining characteristic of the hedgehog ansatz and is
equivalent to the statement that \textit{the field equations reduce
consistently to just one ODE for the profile} $\alpha $.

At this point we can directly evaluate the ansatz in Eq. \eqref{good1} into
Eqs. \eqref{equ1}, \eqref{equ2} and \eqref{equ3} and check whether or not
the same hedgehog property holds. \textit{One could think that the spherical
symmetry is relevant to get a good hedgehog ansatz but the present analysis
show that the hedgehog property does not need spherical symmetry at all}.
Considering that our ansatz satisfies the useful relations in Eqs. %
\eqref{Appconds} and \eqref{good2}, namely 
\begin{equation*}
\nabla _{\mu }\Phi \nabla ^{\mu }\alpha =\nabla _{\mu }\alpha \nabla ^{\mu
}\Theta =\nabla _{\mu }\Phi \nabla ^{\mu }\Phi =\nabla _{\mu }\Theta \nabla
^{\mu }\Phi =\Box \Theta =\Box \Phi =0\ ,
\end{equation*}%
one can see that Eqs. \eqref{equ2} and \eqref{equ3} are identically
satisfied, while Eq. \eqref{equ1} leads to 
\begin{equation}  \label{alpharaw}
-\Box \alpha +\sin (\alpha )\cos (\alpha )\nabla _{\mu }\Theta \nabla ^{\mu
}\Theta +\lambda \sin (\alpha )\cos (\alpha )\nabla _{\mu }\alpha \nabla
^{\mu }\alpha \nabla _{\nu }\Theta \nabla ^{\nu }\Theta =0\ .
\end{equation}%
Eq. \eqref{alpharaw} is an integrable ODE for the soliton profile $\alpha $.
In fact, taking into account that 
\begin{equation*}
\Box \alpha =\frac{1}{L^{2}}\alpha ^{\prime \prime }\ ,\quad \nabla _{\mu
}\Theta \nabla ^{\mu }\Theta =\frac{q^{2}}{L^{2}}\ ,\quad \nabla _{\mu
}\alpha \nabla ^{\mu }\alpha =\frac{1}{L^{2}}\alpha ^{\prime 2}\ ,
\end{equation*}%
the above equation becomes 
\begin{equation*}
\alpha ^{\prime \prime }-\frac{q^{2}}{2}\sin (2\alpha )-\frac{\lambda q^{2}}{%
2L^{2}}\sin (2\alpha )\alpha ^{\prime 2}\ =\ 0\ .
\end{equation*}%
Note also that using 
\begin{equation*}
\sin (2\alpha )\alpha ^{\prime 2}=[\sin ^{2}(\alpha )\alpha ^{\prime
}]^{\prime 2}(\alpha )-\sin^2(\alpha)\alpha ^{\prime \prime } \ ,
\end{equation*}%
we finally arrive to the following equation for $\alpha $: 
\begin{equation*}
\alpha ^{\prime \prime }-\frac{q^{2}}{2}\sin (2\alpha )+\frac{\lambda q^{2}}{%
L^{2}}\sin (\alpha )[\cos (\alpha )\alpha ^{\prime 2}+\sin (\alpha )\alpha
^{\prime \prime }]\ =\ 0\ ,
\end{equation*}%
which, of course, coincides with Eq. \eqref{sessea4}.

\vspace{.5cm}

On the other hand, one could be dissatisfied with the above method to
establish the hedgehog property since in most textbook such a property is
defined by looking at the SU(2) valued field without using the explicit \
parametrization in terms of the fields $\alpha $, $\Theta $ and $\Phi $.
Here we offer another method to derive the hedgehog property which perhaps
is more familiar to many of the readers. This method uses the properties of
the normalized Isospin vector $n_{i}$ of the generalized hedgehog ansatz in
Eq. \eqref{equ1}, \eqref{equ2} and \eqref{equ3}, is using.

Let us remind that the most general parametrization for the Skyrme field is 
\begin{gather*}
U=Y^{0}\mathbb{I}+Y^{i}t_{i}\ ,\quad (Y^{0})^{2}+Y_{i}Y^{i}=1\ , \\
Y^{0}=\cos (\alpha )\ ,\quad Y^{i}=\sin (\alpha )n^{i}\ ,\quad n_{i}n^{i}=1\
, \\
n^{1}=\cos \Theta \sin \Phi \ ,\quad n^{2}=\sin \Theta \sin \Phi \ ,\quad
n^{3}=\cos \Theta \ ,
\end{gather*}%
and the Maurer-Cartan form reads 
\begin{equation*}
L_{\mu }=U^{-1}\nabla _{\mu }U=L_{\mu }^{i}t_{i}\ ,\quad t_{i}=i\sigma _{i}\
,
\end{equation*}%
where the generators of the SU(2) group, $t_{i}$, satisfy 
\begin{equation*}
\lbrack t_{i},t_{j}]=-2\epsilon _{ijk}t^{k}\ ,\qquad t_{i}t_{j}=-\delta _{ij}%
\mathbb{I}-\epsilon _{ijk}t^{k}\ .
\end{equation*}%
One can check that, for the ansatz in Eq. \eqref{good1}, the $n_{i}$ vectors
satisfy the following eigenvalue equation 
\begin{equation}
\Box n^{i}\ =\ \Sigma n^{i}\ , \qquad \Sigma =-\frac{q^{2}}{L^{2}}\ .
\label{eigenn}
\end{equation}%
It is worth to note that the original ansatz for the spherical Skyrmion in
Eqs. (\ref{ori1}) and (\ref{ori2}) satisfies a similar property (but with a
different $\Sigma $ which depends explicitly on the radial coordinate: thus,
we can say that in this sense the present generalized hedgehog ansatz is
simpler than the usual spherical hedgehog ansatz). Eq. \eqref{eigenn} is a
very important result since it allows to reduced all the Skyrme system to
just only one equation for the soliton profile thanks to a very nice
factorization property of the complete field equations (such a factorization
is the matrix version of-and completely equivalent to-the property discussed
above which is responsible for the fact that the three Skyrme field
equations (\ref{equ1}), (\ref{equ2}) and (\ref{equ3}) for $\alpha $, $\Theta 
$ and $\Phi $ reduce to just one equation for $\alpha $).

One can directly check that the components of the Maurer-Cartan tensor $%
L_{\mu }$ defined above are 
\begin{align}
L_{\mu }^{k}\ =\ & Y^{0}\nabla _{\mu }Y^{k}-Y^{k}\nabla _{\mu
}Y^{0}+\epsilon ^{ijk}Y_{i}\nabla _{\mu }Y_{j}  \notag \\
\ =\ & n^{k}\nabla _{\mu }\alpha +\frac{1}{2}\sin (2\alpha )\nabla _{\mu
}n^{k}+\epsilon ^{ijk}\sin ^{2}(\alpha )n_{i}\nabla _{\mu }n_{j}\ .
\label{Rexplicit}
\end{align}%
Now, according to Eq. \eqref{sessea0}, the Skyrme equations have the
following general form 
\begin{equation}
\nabla ^{\mu }\left( L_{\mu }+\frac{\lambda }{4}[L^{\nu },G_{\mu \nu
}]\right) =0\ .  \label{EqSkyrme}
\end{equation}%
Using Eqs. \eqref{eigenn}, \eqref{Rexplicit} and 
\begin{equation*}
\nabla _{\mu }n^{i}\nabla ^{\mu }\alpha =0\ ,\quad n_{i}\nabla _{\mu
}n^{i}=0\ ,\quad \nabla ^{\mu }n_{i}\nabla _{\mu }n^{i}=-\Sigma \ ,
\end{equation*}%
we can compute both terms in the Skyrme equation separately. For the first
term we have 
\begin{equation}
\nabla ^{\mu }L_{\mu }^{k}=n^{k}\left( \Box \alpha +\frac{1}{2}\Sigma \sin
(2\alpha )\right) \ .  \label{dR}
\end{equation}%
Hence, one can see that the divergence $\nabla ^{\mu }L_{\mu }^{k}$ of the $%
L_{\mu }^{k}$ tensor in Eq. \eqref{dR} (that corresponds to the NLSM field
equations) is factorized into the Isospin vector $n^{k}$\ itself (which
obviously never vanishes) time a factor which depends on $\alpha $.
Consequently, in the NLSM case, such a factor is nothing but the equation
for the profile $\alpha $. Hence, the choice of $\alpha $ and of the Isospin
vector $n^{k}$ in Eq. \eqref{equ1}, \eqref{equ2} and \eqref{equ3} reduces
the three field equations of the NLSM 
\begin{equation*}
\nabla ^{\mu }L_{\mu }^{k}=0
\end{equation*}%
to just 
\begin{equation*}
\Box \alpha +\frac{1}{2}\Sigma \sin (2\alpha )=0\ .
\end{equation*}%
The factorization of the divergence $\nabla ^{\mu }L_{\mu }^{k}$ of the $%
L_{\mu }^{k}$ tensor in Eq. \eqref{dR} is the matrix form of the property
stated here above that the field equations (\ref{equ1}), (\ref{equ2}) and (%
\ref{equ3}) reduce to just one equation for the soliton profile $\alpha $:
however, this \textquotedblleft matrix form" of the hedgehog property can be
more familiar to most of the readers so that, for pedagogical reasons we
have included it here in the present discussion. Once again, we see that the
hedgehog property is not related at all with the spherical symmetry and nice
ansatz can be constructed even at finite density and without spherical
symmetry. Even more, the present non-spherical hedgehog, useful to describe
multi-solitonic solutions at finite Baryon density is actually simpler than
the spherical hedgehog (which describes one Skyrmion since the function $%
\Sigma $ in Eq. (\ref{dR}) is constant, as one can see from Eq. (\ref{eigenn}%
)).

One may wonder whether this nice factorization property survives when the
Skyrme term (and, in fact, also the higher order corrections terms mentioned
in the main text) is included. In order to see that this is indeed the case,
one can proceed as follows. In fact, the commutator in Eq. \eqref{EqSkyrme}
can be written as 
\begin{align}  \label{comm}
[L^\nu,G_{\mu\nu}] = 4(S L^k_\mu-S_\mu^\nu L^k_\nu) t_k \ ,
\end{align}
where we have defined 
\begin{align*}
S_\mu^\nu=L_\mu^i L^\nu_i=\nabla_\mu \alpha \nabla^\nu \alpha+\sin^2(\alpha)
\nabla_\mu n^b\nabla^\nu n_b\ ,
\end{align*}
so that 
\begin{align*}
S=L_\mu^i L^\mu_i= \nabla_\mu \alpha \nabla^\mu \alpha-\Sigma \sin^2(\alpha)
\ .
\end{align*}
Using the above relations it is possible to verify that for the second term
in Eq. \eqref{EqSkyrme} we obtain 
\begin{align}
\nabla^\mu (S L^k_\mu-S_\mu^\nu L^k_\nu) = & -\biggl( \frac{1}{2}
\sin(2\alpha)(\nabla_\mu \alpha \nabla^\mu \alpha) +\sin^2(\alpha)\Box\alpha
+\frac{1}{2}\Sigma\sin{(2\alpha)}\sin^2(\alpha)\biggl)\Sigma n^k  \notag \\
& +\frac{1}{2}\sin(2\alpha) \sin^2\alpha \biggl(\nabla^\mu\nabla^\nu
n^k\nabla_\nu n^b \nabla_\mu n_b+\nabla^\mu n^k\nabla^\nu\nabla_\mu
n^b\nabla_\nu n_b\biggl)  \notag \\
& -\sin^4\alpha\varepsilon^{kcd}n_c\biggl(\nabla^\mu\nabla^\nu n^d\nabla_\nu
n^b\nabla_\mu n_b+\nabla^\mu n_d\nabla_\mu\nabla^\nu n^b\nabla_\nu n_b%
\biggl) \ .  \label{Skterm}
\end{align}
Finally, and since for our ansatz we have that 
\begin{gather*}
\nabla^\mu\nabla_\nu n^a\nabla_\mu n^b\nabla^\nu n_b=-n^a \Sigma^2 \ ,
\qquad \nabla^\nu\nabla_\mu n^b\nabla_\nu n_b=0\ ,
\end{gather*}
the Skyrme term in Eq. \eqref{Skterm} takes the form 
\begin{gather}  \label{dcomm}
\nabla^\mu (S L^k_\mu-S_\mu^\nu L^k_\nu)=-\biggl( \frac{1}{2}
\sin(2\alpha)(\nabla_\mu \alpha \nabla^\mu \alpha) +\sin^2(\alpha)\Box\alpha %
\biggl)\Sigma n^k \ ,
\end{gather}
which is also proportional to the vectors $n^k$, as expected. Combining Eqs. %
\eqref{EqSkyrme}, \eqref{dR} and \eqref{dcomm} we obtain again the equation
for $\alpha$ in Eq. \eqref{sessea4}.

This analysis clearly shows that the ansatz defined in Eqs. \eqref{Minkowski}%
, \eqref{sessea1}, \eqref{sessea2} and \eqref{good1} reduces the Skyrme
equations to a single equation for the soliton profile thanks to the
factorization property mentioned here above. It is straightforward to show
that the same derivation is still valid when the higher order terms of the
generalized Skyrme model are included. To the best of authors knowledge,
this is the first complete discussion of the equivalence of these two
different viewpoints on the hedgehog ansatz.

\end{document}